\documentclass[final,5p,times,twocolumn]{elsarticle}
\usepackage{graphicx}
\usepackage{amsmath}
\pdfoutput=1
\journal{Physica D}

\begin{document}
\begin{frontmatter}

\title{\bf About the oscillatory possibilities of the dynamical systems}

\author[upc]{R. Herrero}
\ead{ramon.herrero@upc.edu}
\author[uab]{F. Pi}
\author[uab]{J. Rius}
\author[uab]{G. Orriols\corref{cor1}}
\ead{gaspar.orriols@uab.cat}

\address[upc]{Departament de F\'{i}sica i Enginyeria Nuclear,
Universitat Polit\`{e}cnica de Catalunya, Colom 11, 08222 Terrassa, Spain} 
\address[uab]{Departament de F\'{i}sica, 
Universitat Aut\`{o}noma de Barcelona, 08193 Cerdanyola del Vall\`{e}s, Spain} 
\cortext[cor1]{Corresponding author}

\begin{abstract}
This paper may be ultimately described as an attempt to make feasible the evolutionary emergence of novelty in a supposedly deterministic world whose behavior is associated with that of the mathematical dynamical systems. It means philosophical implications that the paper needs to address, subsidiarily at least. The work was motivated by the observation of complex oscillatory behaviors in a family of physical devices and related mathematical models, for which there is no known explanation in the mainstream of nonlinear dynamics. The paper begins by describing a nonlinear mechanism of oscillatory mode mixing explaining such behaviors and, through its generalization to richer nonlinear vector fields, establishes a generic dynamical scenario with extraordinary oscillatory possibilities, including expansive growing scalability toward high dimensionalities and through nonlinear multiplicities. The scenario is then used to tentatively explain complex oscillatory behaviors observed in nature like those of turbulent fluids and living brains. Finally, by considering the scenario as a dynamic substrate underlying generic aspects of both the functioning and the genesis of complex behaviors in a supposedly deterministic world, a theoretical framework covering the evolutionary development of structural transformations in the time evolution of that world is built up. The analysis includes attempts to clarify the roles of items often invoked apropos of pathways to complexity like chaos, pattern formation, externally-driven bifurcations, hysteresis, irreversibility, and order through random fluctuations. Thermodynamics, as the exclusive field of physics in providing generic evolutionary criteria, is briefly and synthetically considered from the dynamical systems point of view by trying to elucidate its explanatory possibilities concerning the emergence of complexity. Quantum mechanics gets involved in two different ways: the lack of a dynamical systems perspective in the currently accepted interpretations of that fundamental theory and the indeterminacy issues, and both questions are discussed to point out their consequences. The reported evolutionary framework is far from a complete theory but includes both the elements and the skeleton for its tentative building within feasible philosophical grounds. In the lack of alternatives, one should imagine how could be one of such theories and how it could be built, in order to evaluate our approach. In particular, notice that our approach is to a theory of nothing of the physical world but of the underlying reasons for its ordered and creative functioning, which we interpret independent of that world, i.e., a theory of what the Catalan expression "l'entrellat del m\'{o}n" describes so well.
\end{abstract}

\end{frontmatter}

\section{INTRODUCTION}

Every observable phenomenon may be tentatively seen as consisting in things varying in time owing to causal interactions of ones with others and, then, its analysis includes always a problem of dynamics. The structure of causal influences among interrelated things often includes feedback, competition and nonlinearity, the three basic ingredients for a rich dynamical behavior. For this reason, nonlinear dynamics has become one of the currently invoked approaches for trying to understand the occurrence of complex behaviors within the (today apparently naive but commonly adopted) deterministic view of the world \cite{1}. Nonetheless, the intuitively convincing idea that complexity emerges through the participation of an increasing number of degrees of freedom remains away from the dominant standpoints in nonlinear dynamics. After decades of intensive research efforts, the complexity paradigm of nonlinear dynamics remains still focused on chaos, the essence of which is low-dimensional, and there is a lack of knowledge about basic mechanisms yielding high-dimensional phenomena associated with the dynamically organized interplay of a high number of degrees of freedom.
 
Insights on high-dimensionality have been sought by considering sets of coupled nonlinear elements, usually mathematical equations, with a wide variety of  configurations. In these systems, complexity is always looked for in the relative behavior of the coupled units and a rich phenomenology has been found in the form of synchronization, clustering and other static or dynamic patterns \cite{2,3,4}. Similar kinds of effects have been observed in continuously extended nonlinear systems where initially uniform spatial parts show ability to behave differently ones relative to others, due to internal interactions sustained by boundary constraints \cite{5}. Nevertheless, although pattern formation denotes high-dimensionality in the system on its own, the studied dynamical phenomena are usually based on low-dimensional mechanisms affecting the different spatial parts in the proper relative way to yield the observed spatial pattern. For instance, the formation of any kind of static pattern can be associated with one-dimensional processes of the underlying dynamical system \footnote{From the standing eddies in moving fluids to the variety of regular patterns observed in properly bounded media, like for example the Taylor rotating vortices or B\'{e}nard convective cells. More complex, perhaps irregular, but static patterns may also appear under not so proper system boundaries.}, while the K\'{a}rm\'{a}n vortex street in some fluid flows and the rotating spirals in reaction-diffusion systems arise from two-dimensional oscillatory instabilities \cite{7,8,9}. The appearance of more complex spatio-temporal dynamics, like quasiperiodic and chaotic motions \cite{10}, is well understood as far as the processes remain low dimensional. The so-called spatio-temporal chaos \cite{5}, in which decorrelation in both time and space is accompanied by a large number of positive Lyapunov exponents \cite{11,12}, is a clear indication of high-dimensional processes but, as it happens with turbulence, its connection to the known mechanisms of nonlinear dynamics remains unclear. Something similar happens with the variety of behaviors gathered under the name of chaotic itinerancy \cite{CIKiT}. A distinctive feature of the transition routes to complex high-dimensional behaviors is that they occur abruptly, without distinction of the hypothetically accumulated phenomena, thereby making it difficult their association with any sequence of successive bifurcations.

An alternative view for analyzing the emergence of high-dimensional dynamics in deterministic systems is to consider the given system as a whole and to realize that, in general, each dynamical process must affect all of its variables to a more or less extent. Under this perspective the emergence of complex behaviors must manifest itself in the time evolution of each one of the variables and it seems clear that the unique way for this to happen is the successive incorporation of oscillatory modes associated with the dynamical activity of additional degrees of freedom. This was the point of view of Landau in his proposal for tentatively explaining the onset of turbulence in moving fluids on the basis of a quasiperiodic sequence \cite{13}, as well as that of Hopf when presenting a mathematical example displaying features of turbulence \cite{14}, but it is noteworthy that this view is practically absent in the current research of nonlinear dynamics. The main reasons for this abandonment have been the lack of dissipative systems exhibiting long enough quasiperiodic sequences and the usual occurrence of chaos after just a few oscillatory instabilities \cite{15}. These facts, together with the dominant belief that the quasiperiodic motion is the unique way for combining oscillations in nonlinear dynamics, have conferred the exclusivity of complex oscillations on the irregularity of chaos, and this point of view is consequently influencing the attempts for understanding natural systems exhibiting complex (in fact, undecipherable) oscillatory activity like, for instance, the cases of turbulent fluids and living brains.

In this paper we argue against such a widespread opinion by showing that there are robust and generic dynamical mechanisms through which complex time waveforms based on the nonlinear combination of large numbers of oscillation modes can emerge, and by consequently considering the potential relevance of such a kind of oscillatory scenario in relation to complexity. The paper is firstly devoted to offer an answer to the question of how many different oscillation modes can optimally appear together in the observable time evolution of an appropriate $N$-dimensional dynamical system (Sections 2 and 3, and Appendix A). The analysis is founded on the behavior of a class of systems that, implemented in both experimental devices and mathematical models, are able to exploit fully the instability capabilities of a saddle-node pair of fixed points up to sustain complex time evolutions based on the nonlinear combination of $N-1$ different oscillation modes, as previously reported \cite{16,17}. The analysis is developed by considering the extension of the full instability behavior to systems having more general sets of fixed points and the conditions for achieving optimum mode mixing are discussed on generic grounds. We consider a phase space scenario, that we call {\em generalized Landau scenario}, in which the oscillation modes emerged through successive Hopf bifurcations of various fixed points can mix with each other through the intertwinement of invariant manifolds of the corresponding limit cycles. We conjecture that, under optimum circumstances, all of them can appear intermittently together on the time evolution of the same attractor, which not need to be high-dimensional. In addition to the frequency, the characteristic features of a given oscillation mode include a defined oscillatory pattern for each one of the variable properties of the system, in accordance with the phase space orientation of the corresponding limit cycle. Therefore, the trajectories of the mode mixing scenario will describe complex but strictly ordered sequences of intermittent oscillatory patterns with a high degree of potential variability. The phase space topological constraints on the intertwinement of invariant manifolds underlie the mode mixing possibilities and, in this way, prefigure the permitted oscillatory patterns and provide robustness and scalable growing capability for the mode mixing scenario.

In our view, this scenario describes the optimum oscillatory possibilities of $N$-dimensional systems, with N arbitrarily large. We find reason to analyze its potential implication in the real world things and, impelled by apparent parallelisms with the known oscillatory features of turbulent flows and living brains, we have considered each one of these cases by developing conceptual frameworks that, in our view, provide feasibility to the oscillatory scenario involvement (Appendixes B and C). In fact, oscillations, cycles and rhythms are ubiquitously observed in scientific domains covering all the spatial scales, usually in relation to complex systems, and this fact suggests analyzing to what extent the oscillatory scenario could provide a dynamic substrate underlying generic aspects of both the functioning and the genesis of world complexity. This is related to the old philosophical problem of the emergence of novelty in the natural course of the world workings and, in particular, to those of its tentative answers invoking self-organization \cite{18,19,20,21}. The term of self-organization appeared in the context of cybernetics \cite{22} but, even without fitting well into the mainstream of biological thought, its meaning was already implicit in the D'Arcy Thompson's analysis of morphogenesis \cite{23} and it has been tentatively associated with the origin and functioning of life and with the development of biological species \cite{20,21,24}. In these contexts  self-organization mixes with reproduction capability and with adaptation and selection issues, and the evolved system success in proliferating is considered essential for the evolutionary machinery. A wide approach covering all these points has been developed by Kauffman \cite{21} under the guide of the behavior of certain discrete mathematical systems, the so-called random Boolean networks. This approach associates the emergence of order with dynamical scenarios near the edge of chaos by attributing the good working of selection toward achieving fitness to peculiar features of such dynamic regimes. Most notably, by restricting the actual and potential possibilities of the evolving system to those generically compatible with the presumed dynamical scenarios, the approach attributes definability to the underlying dynamical mechanisms at the expense of the adaptation/selection exclusive roles typically assumed in the orthodox views of evolutionary biology. Nevertheless, by associating the concept of structural evolution of a system to the transformation of its dynamically-relevant properties and of their causal interrelations, such a kind of evolution can be appreciated as a component of a wide variety of observable time evolutions in the world, from cosmological to atomic levels, in contexts where both reproduction and adaptation often lack any meaning. It becomes then reasonable to follow the old views in natural philosophy that consider the working of a generic evolutionary motor as an intrinsic feature of the time evolution of the things of the world, with the definability of each one of the evolutionary steps already contained in the details of its own occurrence, independently of subsequent effects like any ulterior evolutionary step or, more globally, the succession of effects hypothetically determining the proliferation success of the evolved system. Along this line and by developing the analysis within the abstract level of the dynamical systems, under the view that their possibilities toward complex behaviors refer to the described oscillatory scenario, we have established a set of elements that, if properly developed, could form a theory for explaining the natural emergence of dynamical organization in the time evolution of a supposedly deterministic world, in intimate association with the underlying reasons for its ordered functioning  (Sect. 4). Concerning living systems, the kind of evolution we are considering is not that of the biological species but that part of the current (time evolution) functioning of life implying modification (in fact, increase) of its underlying structural organization. This includes from the workings of a cell to the learning of a brain, as well as the development of a fertilized egg, and has to do with the problems of the origin of life and of its innovatory capability for sustaining the biological evolution under the sieving stress of selection. We tentatively consider the occurrence of such a structural component of the time evolution of the things of the world as associated with a rather general mechanism, which our framework tries to capture in its essences, and which should cover a wide variety of situations including the most elementary steps of the matter aggregation processes. The framework is rather far away from the nowadays dominant views in the research on complex systems \cite{Nic} and we omit specific comparisons with them.

Among other peculiarities, the paper has no section of conclusions because, in all the considered fronts, the analysis develops under premises and conjectures of pending corroboration and because a number of fundamental problems remain open. A way for, perhaps audaciously, but, in our view, concisely expressing the potential relevance of the paper could be to say that the completion of the theoretical framework, if successful, would imply the transferring connection of nonlinear dynamics from the applied mathematics to the natural science and would provide an alternative approach to that of thermodynamics for tentatively explaining generic evolutionary traits of the world. 

\section{OSCILLATION MODES OF A DYNAMICAL SYSTEM}

The question of how many characteristic frequencies can appear together in the time evolution of an $N$-dimensional system looks basic and simple but, oddly enough, its answer is still pending. According to the predominant linear viewpoint of physicists, the number of oscillation modes that $N$ degrees of freedom can sustain seems limited to $N/2$, i.e., the normal modes of theoretical mechanics. \footnote{In theoretical mechanics one degree of freedom is associated with a pair of conjugate variables that in the general view of nonlinear dynamics are usually considered as two degrees of freedom.} Each mode requires two degrees of freedom and every degree of freedom can work only once in a linear world. By contrast, in the nonlinear dynamics context, the strong influence of chaos makes often plausible the idea of an unlimited number of modes provided $N$ is higher than two, as suggested by the continuous Fourier spectrum of any chaotic signal. But the Fourier components of a time signal need not represent characteristic frequencies of the system and their infinite number in a system of finite dimension points out that there is no physically meaningful decomposition into modes \cite{26}. As a matter of fact, the known routes to chaos do not include processes introducing such a large number of oscillatory modes and we have no reason to associate the non-stable periodic orbits coexisting with the chaotic attractor to different characteristic frequencies of the system. Thus, the broad Fourier spectrum of chaos has no relation with the answer to our question, which may now be reformulated as follows: through which mechanisms a dynamical system can incorporate additional oscillatory modes and how can it combine such oscillations to produce complex time evolutions.

\subsection{Appearance of oscillatory modes}

The theory of bifurcations suggests that the exclusive way for introducing characteristic frequencies into the time dynamics is through the variety of two-dimensional oscillatory instabilities, i.e., the Poincar\'{e}-Andronov-Hopf bifurcation of a fixed point, the Neimark-Sacker bifurcation or secondary Hopf bifurcation of a limit cycle, and the successive bifurcations originating higher-order invariant tori \cite{27,28}. Other bifurcation processes involving periodic orbits, like the period-doubling, cyclic saddle-node and homoclinic or heteroclinic bifurcations, do not contain intrinsic (two-dimensional) mechanisms for the definition of a new frequency, and they must be more properly considered as producing transformation or destruction rather than creation of characteristic oscillatory modes.

Our problem is clearly connected with the physical mechanism suggested by Landau for tentatively explaining the emergence of turbulence in fluids on the basis of a sequence of oscillatory instabilities occurring as the Reynolds number increases \cite{13}. Starting from the stationary laminar flow, successive oscillatory motions superpose ones with others to yield a quasiperiodic evolution with frequencies determined by the boundary conditions and with arbitrary relative phases determined by the moment when a given oscillation begins with respect to the previous one. The role of nonlinearities in this scenario is just to stabilize the oscillatory motion arising from each instability, but they do not participate in the combination of oscillations. In light of the bifurcation theory of finite-dimensional systems, the Landau sequence would correspond to the Hopf bifurcation of a fixed point living in an $N$-dimensional phase space, followed by secondary bifurcations that generate invariant tori of successively higher dimension up to $N/2$. As it is well known, independently of how large $N$ could be, the quasiperiodic sequence cannot be considered a route to chaos because it does not produce sensitivity to initial conditions \cite{29}.
 
The bifurcation theory points also out relations among the different kinds of bifurcations by means of bifurcations of higher codimension. Particularly relevant to our purpose is that the frequency introduced by a secondary or higher-order Hopf bifurcation is often related to the frequency of other Hopf bifurcation of the same fixed point experiencing the primary Hopf bifurcation. This applies to the series of p-torus bifurcations with $1<p\leq{q} \leq{N/2}$, that emerge in the space of dynamical systems from the codimension-q bifurcation associated with the 2q-dimensional eigenvalue degeneracy $\left\{\pm張\omega_{1}, ...., \pm張\omega_{q}\right\}$ of a given fixed point.\footnote{This may be deduced by generalizing results for $q=2$ obtained by means of the method of universal unfoldings \cite{27}.} The Landau sequence would correspond to this series of torus bifurcations and, therefore, the resulting quasiperiodic motion should be based on frequencies related to those of the possible Hopf bifurcations of the initial laminar state, i.e., the unique fixed point implicitly considered in the Landau picture. Something similar happens in the case of integrable Hamiltonian systems, in which the fundamental frequencies are related to the frequencies of the multidimensional toroidal center around the equilibrium state. In this case, the oscillatory motions contained in the flow around the fixed point constitute the so-called normal modes of the (linearized) dynamical system, they appear combined in the quasiperiodic motions with slightly modified frequencies by nonlinear effects, and its maximum number of $N/2$ corresponds to the limit of two-dimensional instabilities that an $N$-dimensional fixed point can sustain. The root of this situation is just the linear viewpoint about the oscillatory possibilities of the dynamical systems.

It is as a matter of fact, however, that the Landau sequence is far from being typical and high-order tori are rarely observed in dissipative systems. This fact may be attributed to an increasing structural fragility with the torus order, as a consequence of homoclinic intersections of the saddle periodic orbits living on the torus when it becomes nonlinearly locked \cite{TOR}, and it may be related to the occurrence of strange attractors when multiple periodic flows on three- or higher-dimensional tori are perturbed \cite{15}. Nevertheless, three-frequency quasiperiodicity has been demonstrated numerically \cite{33} and experimentally \cite{34}, and higher-order tori cannot be excluded although presumably restricted to very small parameter space regions \cite{TOR}. In principle, q-order tori may be expected to be found for systems near enough the codimension-q point, as would be the case for q coupled self-oscillators with weak enough coupling and with each element near its Hopf bifurcation.

On the other hand, it is clear from topological considerations that $N$ dimensions can contain up to $(N-1)$-dimensional tori and the theory of bifurcations provides mechanisms for producing such tori in dynamical systems. In effect, the analysis of universal unfoldings shows that, in certain cases, a two-torus bifurcation can emerge from the three-dimensional eigenvalue degeneracy $\left\{0,\pm張\omega\right\}$ and a three-torus bifurcation from the four-dimensional degeneracy $\left\{\pm張\omega_{1},\pm張\omega_{2}\right\}$ \cite{27}. In these bifurcations the new frequency does not appear related to any Hopf bifurcation of the fixed points and, beginning from zero, its value grows with a parametric distance to the eigenvalue degeneracy.\footnote{Like happens with the Hopf bifurcation of a fixed point with respect to the Takens-Bodganov point of the two-dimensional degeneracy $\left\{0,0\right\}$.} These bifurcations represent therefore a way for the emergence of additional characteristic frequencies, but it is noteworthy that the corresponding tori seem still more delicate than those related to the Hopf bifurcations of a fixed point. The reason may be that, in addition to their intrinsic structural fragility, these tori can also be affected by heteroclinic connections of previously existing saddle sets \cite{27} and, as a matter of fact, they have only been numerically observed in extremely small regions of the parameter space \cite{36,37,38}.
 
In this way we find reason to conclude that the emergence of oscillatory modes in a dissipative dynamical system is actually associated with the ability of its fixed points to undergo Hopf bifurcations and that the coexistence of fixed points could be the first step towards developing complex oscillatory behaviors.\footnote{The various fixed points can experience Hopf bifurcations independently of the state where the system actually is. At this respect, we will use the term oscillatory instability to refer not just to the instabilities actually experienced by the system but to those occurred everywhere in its phase space.} The next relevant question is to what extent the oscillation modes emerged from either the same or different fixed points can mix ones with others to appear together on the observable time evolution of the system.

\subsection{Nonlinear mode mixing}

Contrary to common beliefs, the quasiperiodic motion does not represent the exclusive way for combining oscillations in dissipative nonlinear systems. There is another type of mixing mechanism that, in the presence of dissipation, appears to be more robust and perhaps more basic than the creation of invariant tori. The mechanism is essentially and simply based on the intertwine of the oscillatory motions emerged in definite phase space regions in association with the Hopf bifurcations of one or more fixed points, and it develops without necessarily requiring the creation of limit sets of successively higher dimension. Under proper conditions, the mixing becomes apparent in the time evolution of an attractive limit set in the form of intermittent bursts of the different oscillation modes, combined ones over the others according to the scale of frequencies, and it occurs because the attracting set extends into the corresponding phase space regions \cite{16,17}.

\begin{figure}[tbp]
\centering
\includegraphics[width=0.6\linewidth]{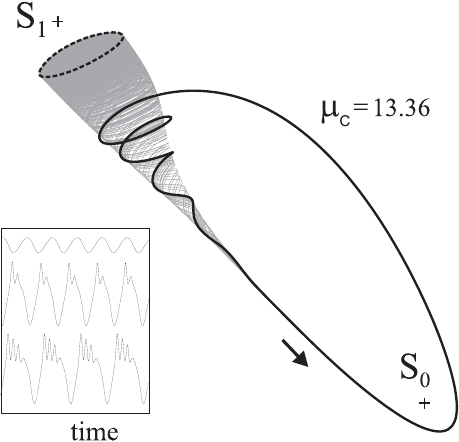}
\caption{Phase space representation of numerical results for an $N=3$ system illustrating the nonlinear mixing of two oscillation modes emerged from a saddle-node pair of fixed points. The Hopf bifurcations take place at $\mu_{C}$=12.96 on the saddle $S_{1}$ and at 13.10 on the node $S_{0}$. The represented stable orbit has grown with $\mu_{C}$ incorporating helical turns around the unstable manifold (in grey) of the saddle cycle (broken), without undergoing any bifurcation and remaining strictly periodic. The inset shows the time evolution of one of the variables ($\psi$) for three control parameter values: from just after the bifurcation up to the represented attractor.}\label{Fig1}
\end{figure}

\begin{figure}[tbp]
\centering
\includegraphics[width=0.8\linewidth]{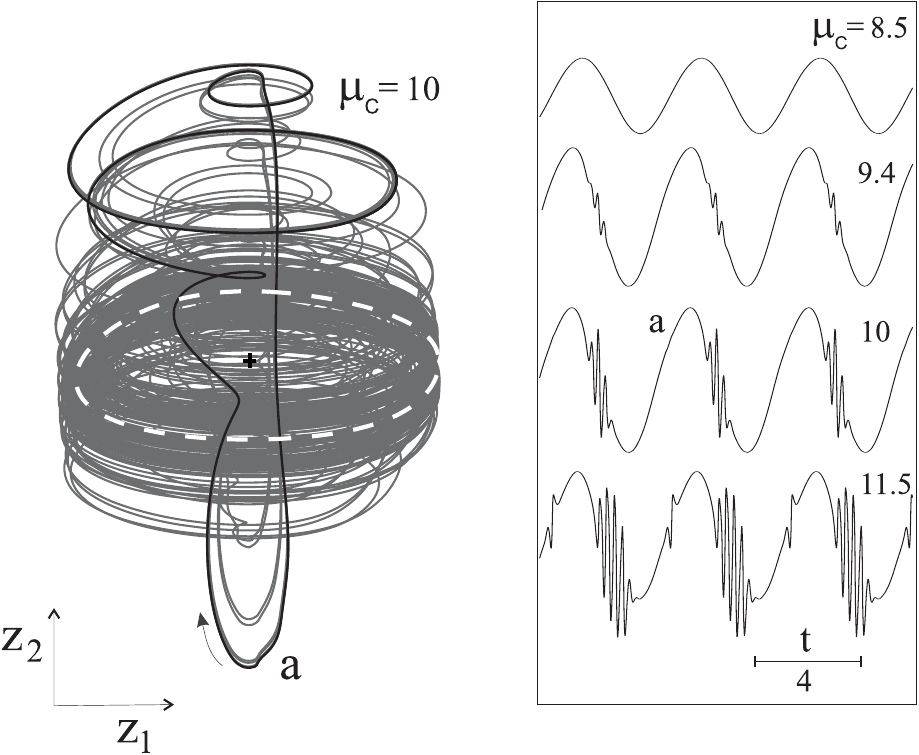}
\caption{Numerical results for $N=4$ illustrating how two oscillation modes emerged in successive Hopf bifurcations of the same fixed point mix without a torus bifurcation. The Hopf bifurcations take place at $\mu_{C}$=8.2 and 9.1 with angular frequencies of 1.41 and 25, respectively. The stable orbit (black) has grown with $\mu_{C}$ incorporating helical turns around the three-dimensional unstable manifold (in grey) of the saddle cycle (broken white), without undergoing any bifurcation up to $\mu_{C}$=12, where a period doubling occurs. Label a denotes where a second helical structure will appear. The unstable manifold is represented through a few of the trajectories to facilitate its visualization. The saddle cycle does a subcritical torus bifurcation at $\mu_{C}$=10.3 with a secondary frequency of 1.38 and it becomes stable within a saddle torus. The inset shows the time evolution of one of the variables ($\psi$) for the stable orbit at different control parameter values.}\label{Fig1bis}
\end{figure}

A well-known example of this type of mode mixing is offered by the distinctive bursting activity of some types of neurons \cite{40}. It corresponds to a generic dynamical behavior that, for systems of dimension three or higher, occurs when a stable limit cycle is growing under parameter variations toward a transversely oriented saddle limit cycle and becomes one of the so-called Shilnikov-type attractors, as that shown in Fig.~\ref{Fig1}. The time evolution of these attractors reflects the nonlinear mixing of two oscillation modes associated with the Hopf bifurcations of two different (but saddle-node connected) fixed points.\footnote{In neural models, bursting is traditionally studied with fast/slow decomposition approaches so useful for analytical purposes \cite{41}, but the full phase space perspective provides the most generic view of what a bursting is.} The important feature is that the intermittent incorporation of oscillations at the saddle frequency does not necessarily require any bifurcation of the stable orbit. This can be deduced from the analysis of two-parameter Poincar\'{e} maps describing a three-dimensional flow with a saddle limit cycle near homoclinicity \cite{42,43} and it has been verified in numerical simulations, as in the case of Fig.~\ref{Fig1}. Less well-known is the kind of nonlinear mixing shown in Fig.~\ref{Fig1bis}, in which a initially stable fixed point has experienced two successive Hopf bifurcations and the second oscillation mode (of higher frequency) locally incorporates in the first limit cycle while it remains strictly periodic. Here again the incorporative mixing takes place through the intertwinement with the unstable manifold of the second limit cycle, without requiring the occurrence of any bifurcation, but now the fast oscillations appear at two different places of the first cycle.\footnote{By modifying the system it is possible to achieve the Hopf bifurcation of the fast frequency earlier than that of the lower one. In this case the mode mixing on the atractor usually happens through a torus bifurcation yielding a quasiperiodic orbit of the two frequencies but, with increasing the control parameter, the quasiperiodic signal usually transforms to become a two-sided bursting waveform, probably involving the torus breakdown. So that, in practice, the two ways of mode mixing from the same fixed point would produce equivalent time evolutions, at least when the frequencies are clearly distinct.} Another significant difference is that the period of the stable orbit does not change with the incorporation of fast oscillations while in Fig.~\ref{Fig1} the period increases clearly. This denotes the occurrence of a homoclinic process when the mixed modes arise from different fixed points, as well as its absence when the modes emerge from the same fixed point. Finally, in the represented variable, $\psi$, the two kinds of mode mixing appear additionally differentiated by the fact that in one kind the fast oscillations emerge on the top (or the bottom, according to the relative position of the saddle fixed point) while in the other one they emerge at intermediate levels of the oscillatory undulation. All these features should be taken into account when analyzing more complex oscillatory waveforms because the simultaneous mixing of a number of oscillations emerged from a saddle-node pair of fixed points is a combination of the two described mechanisms of nonlinear mode mixing.

\begin{figure}[tbp]
\centering
\includegraphics[width=0.9\linewidth]{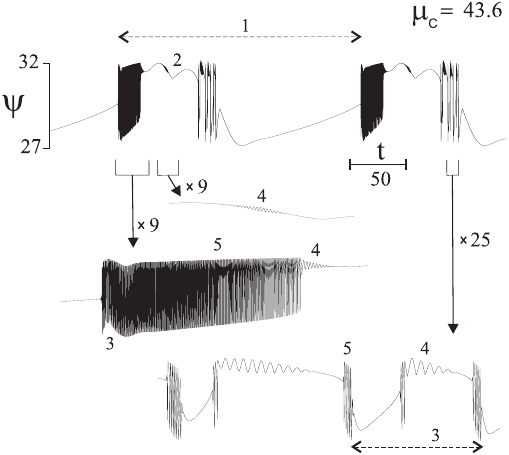}
\caption{Numerical results showing the nonlinear mixing of 5 oscillation modes in the time evolution of an $N=6$ system for the given value of the control parameter $\mu_{C}$. The system of equations (2)-(4) has been designed (see Subsection 3.2) by imposing Hopf bifurcations of angular frequencies 0.04, 0.25, 2, 20, and 125, alternatively in a saddle-node pair of fixed points and according to the values of $p$ equal to -6, 15.6, -7.4, 16.2, and -6.9, and by choosing $c_{1}= 250$. The nonlinear function is $g(\psi)=(1.25-1.06\cos\psi)/(1.68-\cos\psi)$, with which the Hopf bifurcations on the involved saddle-node pair of fixed points happen at $\mu_{C}$= 38.5, 53.5, 40.0, 54.8, and 39.4, respectively. Numeric labels denote the different modes ordered from lower to higher frequencies. Odd (even) numbers correspond to the node (saddle) point. More than 200 previous cycles have been discarded in the represented signal to assure its approach to asymptotic behavior. The orbit periodicity has not been strictly verified but the successive cycles show extremely similar structure. The oscillation modes 2 and 4 appear on the time evolution even if the corresponding Hopf bifurcations will occur at higher $\mu_{C}$ values, suggesting the subcriticality of such bifurcations.}\label{Fig2}
\end{figure}

Figures 3 and 4 illustrate the nonlinear intertwinement of five oscillatory modes of clearly distinguishable frequencies in the dynamics of a system of dimension 6. In a phase space of dimension $N$, a saddle-node pair of fixed points can sustain up to $N-1$ different Hopf bifurcations and the system of Fig.~\ref{Fig2} is able to exploit such possibilities fully. Throughout the course of parameter variations, $N-1$ limit cycles have probably emerged or are close to appear and some invariant tori could have been created. The cluster of limit sets contains one attractor and a variety of saddles with the common feature of having a branch of their unstable manifold ending toward the attractor. Complex secondary processes can occur but they produce 
nonlinear mode mixing with relatively generic features. In essence, what happens is that the attractor incorporates localized helical motions related to the influence of neighboring saddles and, in this way, the observed time dynamics describes a complex combination of oscillatory bursts that recurrently repeats. See details in the caption of Fig.~\ref{Fig2} but, to properly distinguish the two basic mechanisms of mode mixing, we need here to remark that modes 1, 3 and 5 correspond to the node fixed point while modes 2 and 4 to the saddle point. The features of mixing between modes emerged from the same fixed point are clearly appreciated in the incorporation of mode 5 in mode 3, of mode 3 in mode 1 and of mode 4 in mode 2. At the left side of mode 1, the incorporation of modes 3 and 5 appears with dominance of the fast mode and then it looks like a temporary quasiperiodic signal, while at the other side a more graded combination happen. The features of mixing between different fixed points are seen in the incorporation of mode 2 in mode 1 and of mode 4 in mode 3. The case of mode 4 is significant because it appears through the two mechanisms on the same attractor and suggests in this way the richness of the mode mixing processes. First, the fact that the incorporation of mode 4 in mode 2 manifests in mode 1, i.e., the attractor, points out the effective working of a transmissive chain of influences. Such a transmission of mode influences is also implied by how the incorporation of mode 4 in mode 3 displays on the attractor but, additionally in this case, it suggests also the occurrence of homoclinic events involving secondary limit cycles emerged from the saddle-node pair of fixed points.

\begin{figure}[tbp]
\centering
\includegraphics[width=0.74\linewidth]{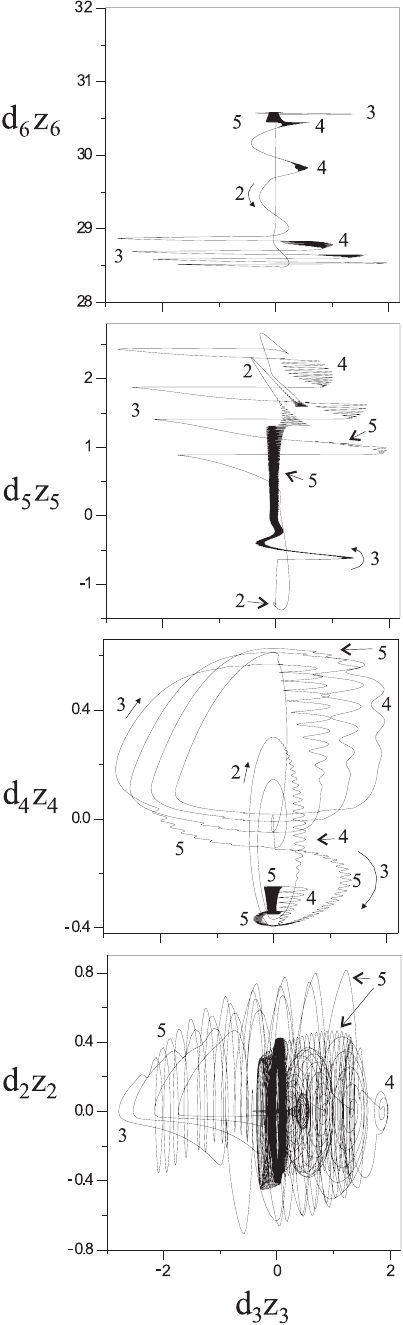}
\caption{The orbit of Fig.~\ref{Fig2} projected in different planes of the six-dimensional phase space. The fixed points are located on the $z_{N}$ axis ($S_{0}$ and $S_{1}$ are at $d_{6}z_{6}$ equal to 29.7 and 31.9) and then the two points appear superposed at the origin of the planes perpendicular to this axis. Numeric labels denote the different oscillatory modes.}\label{Fig3}
\end{figure}

In case of clearly different frequencies, like in Fig.~\ref{Fig2}, the basic regularity associated with the lowest frequency mode often appears as practically periodic, regardless of the extremely complex orbit structure, and this makes a clear distinction with respect to chaos. On the other hand, the intermittent activity of the rest of oscillation modes implies the lack of any frequency-locking and consequent resonance problems. This explains the robustness of the multiple-bursting waveform and it is related to the fact that the underlying attractor development occurs without necessarily requiring high-dimensional limit sets but simply experiencing a continuous transformation associated with the phase space flow. Another remarkable difference of the bursting signal with respect to the quasiperiodic motion is that it cannot be expressed as a linear superposition of the combined oscillatory modes, their harmonics or  oscillations with any linear combination of their frequencies. In other words, the Fourier spectrum does not reflect directly the mode structure of the bursting signal because the intermittent mode combination is strictly nonlinear.

In order to analyze how generic the mixing mechanism may be, we try to imagine dynamical systems with large sets of fixed points potentially able to exploit their oscillatory possibilities and, in this way, the three following questions arise: what sets of fixed points, how many different Hopf bifurcations can each fixed point sustain, and to what extent the oscillation modes can be mixed together.

\section{VECTOR FIELDS WITH A MULTIDIRECTIONAL NONLINEAR PART}

A very general description of the $N$-dimensional systems useful for analyzing the possibilities for the emergence of complex time dynamics is as follows
\begin{equation} \label{equation 1}
\frac{dx}{dt}=Ax+\sum^{m}_{j=1}b_{j}f_{j}\left(x,\mu\right),\\
\end{equation}

\noindent
where $x\in\Re^{N}$ is the vector state, $A$ is a constant $N$x$N$ matrix, $b_{j}$ are constant $N$-vectors, $f_{j}$ are scalar-valued functions nonlinear in $x$, $\mu$ describes constant parameters involved in the nonlinear functions, and the $m\leq N$ components $b_{j}f_{j} $ are assumed linearly independent.\footnote{The decomposition in linear and nonlinear parts can change in a transformation but the directionality degree $m$ should remain invariant in general.}

\begin{figure}[t]
\centering
\includegraphics[width=0.48\linewidth]{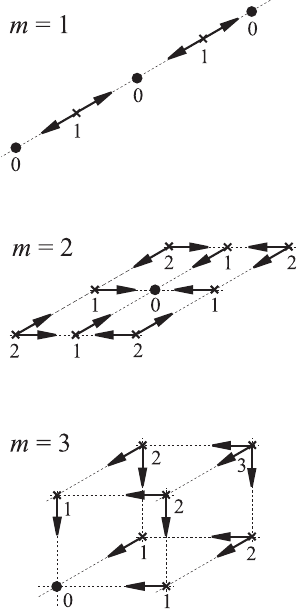}
\caption{Arrays of fixed points within $m$-dimensional linear subspaces of the phase space, for $N$-dimensional systems having efficient nonlinearities in $m$ linearly-independent components of the vector field. Numeric labels denote unstable manifold dimensions of the equilibria by supposing the dimensions outside the $m$-dimensional subspaces to be attractive for physical reasons. Regular arrays are drawn for simplicity but both the separation distance and position alignment of equilibria will be surely irregular within the linear subspaces. Only one attraction basin is represented for $m=2$ and only one of the basin corners is represented for $m=3$. In general, the arrays would extend at the other side of the separatrix with additional attractors.}\label{Fig4}
\end{figure}

The relevant detail to our purpose is that the multi-directionality of the nonlinear part of the vector field determines the topological structure of the potential sets of fixed points in phase space. In effect, with a linear transformation to a proper basis including the vectors $b_{j}$ it may be seen that, in general, the equilibria must be contained within an $m$-dimensional linear subspace determined by the $b_{j}$ and $A$.\footnote{Let be $\left(b_{1},..,b_{m},a_{m+1}, ..,a_{N}\right)$ one of such basis, in which every $a_{q}$ has been chosen to be orthogonal to all the $b_{j}$, then the projections onto the vectors $a_{q}$ of the condition $0=Ax+\sum b_{j}f_{j}\left(x,\mu\right)$ shows that the equilibria should be contained into the $(N-1)$-dimensional hyperplanes passing through the origin: $a_{q}Ax=0,~q=m+1,..,N$. If $A$ is non singular, the normal vectors of such hyperplanes, $a_{q}A$, are linearly independent and the intersection of the $N-m$ hyperplanes reduces to an $m$-dimensional linear subspace. The fixed points actually existing within such a subspace are determined by the $m$ projections of the equilibrium condition onto the vectors $b_{j}$.} With generic considerations and assuming effective enough nonlinear functions we imagine scenarios like those shown in Fig.~\ref{Fig4}, where $m$-dimensional arrays of fixed points have appeared through differently oriented saddle-node bifurcations and, more rarely, through pitchfork bifurcations.\footnote{The pitchfork bifurcation is of codimension-one but requires particular conditions on the nonlinearity that, although not strictly necessary, are usually achieved through a proper symmetry.} Each stable node has the basin of attraction defined by the stable manifolds of the surrounding saddle points and, assuming full arrays, we have a set of $3^{m}-1$ saddle points on the separatrix (see details in Appendix A). 

It is worth stressing the generic features of the picture we are describing in relation to physically relevant situations in the sense that, if the fixed points exist and they are not accompanied by other limit sets, the dimensional structure of stable and unstable manifolds must generically be like that indicated in the figure. To realize this fact, we imagine a proper deformation of system (1) until it has a unique fixed point, which we assume stable for physical reasons; then we gradually modify the system to achieve an arbitrary succession of single zero eigenvalue bifurcations creating pairs of additional fixed points and we generically obtain arrays of fixed points like those shown in the figure. Notice the hierarchy of connections among the equilibria, in the sense that each one of them is saddle-node connected with and only with the $2m$ neighboring points having unstable manifolds differing with it by one dimension. These one-dimensional saddle-node connections mark the ways through which the fixed points can approach one another until they merge and disappear by pairs in single zero eigenvalue bifurcations. And, most importantly for our analysis, such connections constitute the skeleton of the structure of invariant sets through which the nonlinear mixing of oscillation modes should occur.

Let us also remark that the occurrence of multi-dimensional arrays of equilibria is by no means rare because it only requires proper nonlinear functions. Of course, actual arrays will surely be incomplete and a stable node will probably be surrounded by only a few of the $3^{m}-1$ saddles potentially possible on the separatrix, but the exponential growing with $m$ makes situations with a large number of fixed points feasible. A typical situation for achieving $m$-dimensional arrays of fixed points is the case of $m$ weakly coupled oscillators with each element possessing multiple equilibria individually.\footnote{The inherent difficulty of the equilibria analysis is just the reason for the lack of consideration of this basic aspect in the plethora of publications dealing with coupled oscillators. As exceptions see \cite{47}.} Fluid flows provide another example of situations with large arrays of equilibria, each one of them corresponding to a different steady flow structure and almost all of them having unstable dimensions \cite{48,49}.

Concerning the oscillatory possibilities, we begin supposing that a fixed point may sustain successive two-dimensional bifurcations up to exhaust its stable and unstable manifolds one time and realizing that, in mean, this number is $(N-1)/2$, as may be seen by considering any pair of saddle-node connected fixed points together, independently of the even or odd value of $N$. This number expresses the linear possibilities of the system through the given fixed point. The nonlinearities, in addition to sustain the two-dimensional stabilization of each oscillation mode in a limit cycle, should provide for the coexistence of fixed points, for the coexistence of their oscillation modes and for the working of the mode mixing mechanism. While the occurrence of fixed points and their Hopf bifurcations can, in principle, be considered fully attainable by means of a proper system design, the assumption of efficient mode mixing requires some analysis. Starting from our experience with $m=1$ systems \cite{16,17} (see subsection 3.2) and analyzing how the invariant manifolds of the limit cycles may be for $N$-dimensional systems with $m=2$ and 3, we extend the analysis to the general case and arrive at what we call the generalized Landau scenario for the emergence of complex oscillatory behaviors in dynamical systems. The details are given in the Appendix A, where we consider the optimum circumstances for achieving the full instability behavior in $(N,m)$ systems as defined by Eq.(1), while the essence of the reasoning is presented in the next subsection.

\subsection{Generalized Landau scenario}

Let it be the cluster formed by one attractive node and the full set of saddle points located on the separatrix of the basin of attraction. The first Hopf bifurcation of the stable node will probably produce a stable limit cycle, while the succeeding bifurcations of this fixed point and the bifurcations of the saddle points will produce saddle limit cycles of different types, but all of them will have a branch of their unstable manifold ending toward the attracting cycle. Very complex processes will probably occur but we assume as a generic feature the presence of one attractor and a cluster of saddle limit sets within a connecting structure of invariant manifolds related to that of the initial array of fixed points. According to our interpretation, optimum mode mixing possibilities over the attractor are achieved by assuming Hopf bifurcations only within the stable manifolds of the fixed points, while their initially unstable manifolds don't participate to preserve the way of influence toward the attractor.

The schematic drawings of Fig.~\ref{Fig5} illustrate how the attractor emerged from the stable node can receive the influence of the saddle-node connected saddles. The label $S_{q}$ is used to indicate fixed points with unstable manifold of dimension $q$ before any Hopf bifurcation and, for concreteness, the label is maintained after the occurrence of the bifurcations. Thus, the situation of Fig.~\ref{Fig5} describes the influence of saddles of type $S_{1}$ on the attractor emerged from $S_{0}$:

\begin{figure}[t]
\centering
\includegraphics[width=0.38\linewidth]{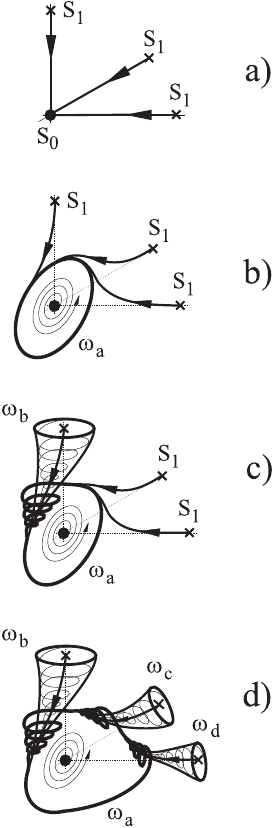}
\caption{Nonlinear mode mixing in a set of $S_{1}$ points saddle-node connected to a given $S_{0}$ point. The scheme shows how the first oscillation modes emerged from different $S_{1}$ become incorporated on the stable orbit appeared from $S_{0}$. In the $N$-dimensional phase space, successive Hopf bifurcations on stable dimensions of $S_{0}$ will produce additional mode mixing on particular zones of the stable orbit and the same for successive Hopf bifurcations on stable dimensions of the $S_{1}$ points (not represented in the figure).}\label{Fig5}
\end{figure}

\begin{figure}[tp]
\centering
\includegraphics[width=0.85\linewidth]{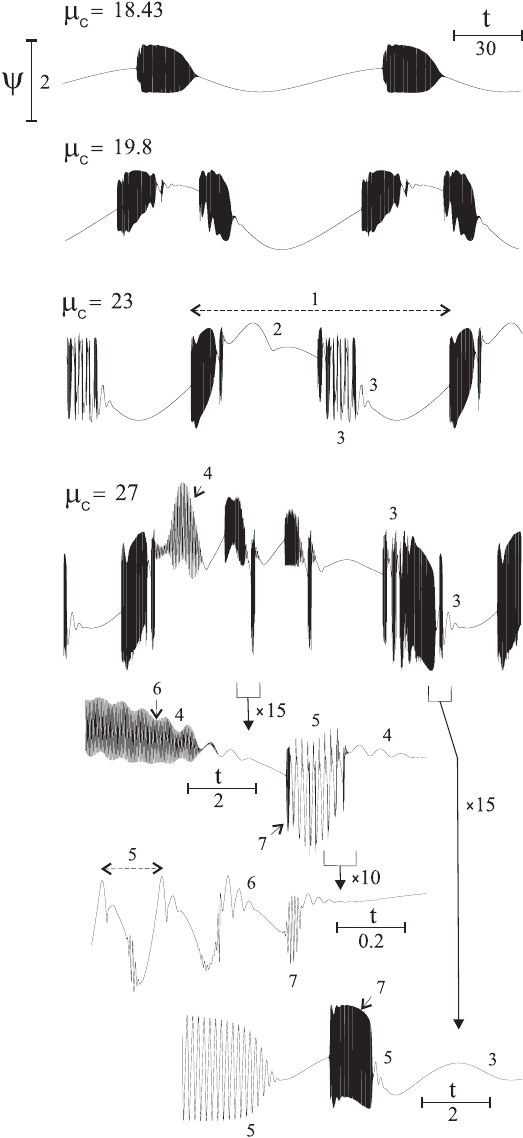}
\caption{Numerical time evolutions showing gradual nonlinear mixing of up to 7 oscillation modes with increasing $\mu_{C}$, for an $N=8$ system designed by imposing Hopf bifurcations of angular frequencies 0.06, 0.3, 2, 10, 40, 200, and 800 at respective $p$ values equal to -6, 7, -7, 8, -6.5, 9, and -7, which alternatively correspond to the node and saddle fixed points, and by choosing $c_{1}$=1000. The nonlinear function is the same as in Fig.~\ref{Fig2}. The four reported signals look practically periodic.}\label{Fig6}
\end{figure}

\begin{itemize}\setlength{\itemsep}{-4pt} \setlength{\topsep}{0pt}
\item[i)] The first Hopf bifurcation of $S_{0}$ originates a stable limit cycle \footnote{Created directly if the Hopf bifurcation is supercritical or in a previous cyclic saddle-node bifurcation if subcritical. We assume proper nonlinearities for sustaining an attractor around the first two-dimensional instability of $S_{0}$.}, to which the endings of the one-dimensional (1D) unstable manifolds of the $S_{1}$ points are transferred (Fig. 6b). On the other hand, $S_{0}$ has incorporated a 2D unstable manifold spiraling toward the stable cycle that will yield, in the next Hopf bifurcation, a 3D unstable manifold of the second limit cycle through which its oscillation mode will be transmitted to the stable cycle (like in Fig.~\ref{Fig1bis}) and so on for the successive bifurcations of $S_{0}$.
\item[ii)] The first Hopf bifurcation of a given $S_{1}$ produces a saddle cycle with a 2D cone-shaped unstable manifold bordering the 3D unstable manifold of the point (Figs. 6c and 1). This structure has emerged through a $2$D expansion of the previous 1D manifold of the bifurcating point and, under its guide, ends with asymptotic tangency on the stable cycle. It contains the essence of the intertwine mechanism of mixing: the flow over and around the cone-shaped manifold is a helical motion at the frequency of the saddle cycle that can reach the stable cycle by influencing its shape and time evolution. The mixing occurs locally in the contact region and it works like a corkscrew during the parametric growing of the stable cycle. The helical turns remain roughly parallel to the saddle cycle so that, in addition to the frequency, the relative effect of the oscillation mode on the system variables is preserved in the mixing process. The corkscrew-like growth of the stable cycle toward the saddle cycle is associated with a homoclinic process at the end of which the growing cycle would disappear by making tangency to the saddle cycle and to its stable manifold. The homoclinic process regulates the efficiency of mixing, i.e., the number of helical turns, but it is worth to stress that mode mixing does not require the fulfillment of homoclinicity.
\item[iii)] Successive instabilities over pairs of stable dimensions of $S_{1}$ produce limit cycles with unstable manifolds of successively higher dimension having, in particular, a 3D submanifold linked to the stable manifold of the previous cycle (similarly to the $S_{0}$ case of Fig.~\ref{Fig1bis}). Is through this submanifold that the emergent mode is incorporated to the previous cycle and through the whole unstable manifold that the new mode is transferred toward the attractor and, possibly, other limit cycles appeared from the node point. For instance, the second cycle has a 4D unstable manifold bordering the 5D unstable manifold of $S_{1}$ and both have emerged as a two-dimensional expansion of the previous 3D unstable manifold of the fixed point. We don't know well how such an expansion really works and how it may affect the unstable manifold of the first cycle. Nevertheless, we can affirm that the 5D structure of unstable manifolds don't remain bounded by the helicoidal turns of the growing stable cycle because this cycle is incompatible with becoming homoclinic to the second cycle of $S_{1}$. This means that the second mode of $S_{1}$ can influence other limit cycles emerged from $S_{0}$, particularly the second one because it is compatible with becoming homoclinic to the second cycle of $S_{1}$, i.e., the second cycle of $S_{0}$ can experience a corkscrew-like effect around an unstable submanifold of the second cycle of $S_{1}$. The mixing mechanism at the level of $S_{0}$ would then transfer to the attractor the second mode of $S_{0}$ with intermittent contributions of the second mode of $S_{1}$ at every undulation.
\footnote{This sequence of mixing mechanisms can explain how mode 4 appears within mode 3 in the attractor of Fig.~\ref{Fig2}, or how mode 6 appears in mode 5 in the attractor of Fig.~\ref{Fig6}. The generic nature of the process is pointed out in Fig. 10 of \cite{17} where the mixings of mode 4 in mode 3, mode 6 in mode 5 and mode 8 in mode 7 appear on the same attractor for an $N=10$ system.}
In addition, the second Hopf bifurcation of $S_{1}$ must affect the 2D unstable manifold of the first cycle in such a way that the influence of the second cycle on the first one should be transferred toward the attractor. This means that the cone-shaped manifold should incorporate an appropriate localized folding introducing the intermittent mixing in the corkscrew effect upon the attractor.
\footnote{This could explain how mode 4 appears within mode 2 in the attractor of Fig.~\ref{Fig2}, or how modes 4 and 6 appear in mode 2 in the attractor of Fig.~\ref{Fig7}.} Generalizing we conclude that all the oscillation modes emerged from $S_{1}$ up to exhaust its stable manifold may appear mixed in the attractor and this can occur simultaneously in a two-fold way: first, through mixing at the level of $S_{1}$ and then falling down through the corkscrew effect around the unstable manifold of the first of the modes and, second, through direct corkscrew influence from upper to lower levels between modes of equal order and then mixing at the level of $S_{0}$.
\item[iv)]	The different $S_{1}$ points connected to $S_{0}$ can originate mixing of their oscillation modes over the same attractor, on the corresponding zones of tangency and independently ones of others (Fig. 6d). The simultaneity of efficient mode mixing for various $S_{1}$ points implies that a heteroclinic process is approaching, at the end of which the growing cycle would disappear by closing a cyclic connection among the involved saddle cycles.
\end{itemize}

The optimum scenario of Appendix A considers the full set of fixed points in the basin of attraction and describes a chain of mode mixing influences from top to bottom in the $j$ scale of $S_{j}$ points that should be able to transmit intermittent contributions of the various oscillation modes emerged within the stable manifolds of all the fixed points toward the attractor. Thus, the conjectured scenario describes a way through which $N$ degrees of freedom might sustain a complex dynamical activity based on the intermittent combination of a large number of oscillation modes and with a multiplicity of combinatory pathways.

Concerning the oscillation frequencies it is clear that the mode mixing mechanisms imply restrictions on their values because a bursting of oscillations should necessarily be of higher frequency than the oscillation into which it is incorporated. This means that the described scenario could gradually transform when choosing more similar Hopf frequency values for a pair of saddle-node connected points. In particular, the time evolution waveforms would surely look different as the various oscillation modes become alike, while the described mixing mechanisms could become less efficient and perhaps different ones could begin to work. The mode mixing efficiency depends also on other factors like, for instance, the vector field divergence determining the dissipative level of the system. \footnote{See Fig. 12 of \cite{16}.}

The mode mixing scenario deserves a more careful analysis to put it in proper context within the theory of nonlinear dynamics. In the meantime, we consider it as a global process affecting the flow of the phase space region where the complex structure of interrelated invariant sets have been or will be created. Parametrically speaking, the process develops with the appearance of fixed points and the occurrence of Hopf bifurcations within the stable manifolds of these points, while the mode mixing influence upon the attractor happens under the guide of the unstable submanifolds of the several saddle limit sets, with more or less efficiency according to the proximity of such manifolds to close homoclinic and heteroclinic loops. On the other hand, in addition to the enhancement of the corkscrew-like mixing, the proximity of saddle connections makes likely the occurrence of complex bifurcation sequences and chaos. Nevertheless, we interpret the mode mixing process as a continuous deformation of the flow occurring independently of such bifurcations. In principle, a simple periodic orbit could intermittently incorporate the total number of oscillation modes, although the parametric accumulation of phase space events would surely imply abrupt changes in the observable attractor. If chaos would occur, it should be an additional effect, with the possible strange attractor and each one of the many non-stable periodic orbits coexisting with it having similarly complex orbit structures based on the intermittent mode combination. Thus, any periodic window will exhibit similar complex waveforms while the chaotic evolution will manifest through irregularities in the structure of successive cycles that, for clearly different frequencies, will typically appear to be slight. Nevertheless, the irregularity degree often enhances when the nonlinearity strength is increased through larger $\mu_{C}$ values, even for clearly different frequencies, and, in general, the less different the oscillation frequencies are choosen, the more likely the presence of chaos becomes. \footnote{This is illustrated in Figs. 9 and 13 of \cite{16}, where the corresponding Lyapunov exponents are reported.} Since the complex waveform structure is not related to stability features of the orbit, not to the dimension of the asymptotic invariant set, the evaluation of Lyapunov exponents and attractor dimensionalities does not provide any characterization of such a complex waveform. A Poincar\'{e} map of the orbit is also ineffective in capturing the intermitent mode mixing structure and this brings us to the relevant and more general question of to what extent the generalized Landau escenario is achievable in discrete systems.

The complex oscillatory behavior would possess robustness, organization and recurrence. There is robustness against parametric variations because of the gradual nature of the mixing process and the lack of mode resonance problems. There is organization in the way the different oscillation modes appear sequentially combined, as intrinsically regulated by the structure of invariant manifolds, and in how such a mode combination is peculiar for each one of the system variables, as determined by the orientation of the various limit cycles in the phase space. The organized structure of oscillation modes displays features like intermittency, self-similarity, redundancy and scalability. And, finally, there is recurrence of the complex dynamical activity at the slowest frequency characteristic mode, as it is obliged in the time evolution of any attractor, and there are also successive levels of transitional recurrences associated with the similarity levels of the evolution.

In the case of systems with spatially distributed dynamical properties, like presumably happens in any real high-dimensional system, the simultaneous observation of a number of local values of one of such properties as a function of time would show the corresponding spatio-temporal projection of the complex oscillatory activity of the system. To achieve a generic visualization of the system behavior in the physical space, we need to imagine how the phase space entities look in that space: what are the fixed points, what are the various limit cycles of different orientation emerged from a given fixed point, and what represents the intertwine mixing of the oscillation modes upon the observable attractor.\footnote{To remain within a finite-dimensional perspective, one can generically imagine a cellular decomposition of the spatial region occupied by the system and associate different phase space coordinates with scalar physical properties of the various cells, instead of considering the appropriate function space. The dimension would be the number of relevant scalar properties times the number of cells.} Even assuming fixed points with static spatial patterns of good contrast and regular form, the spatial structure of the oscillatory state may be expected to become quickly obscured as the number of mixed modes increases.

Finally, it is worth considering the feasibility of systems exhibiting so high degrees of oscillatory instability behavior. For the case of systems with $m=1$, we have been able to design experimental devices \cite{17} and $N$-dimensional mathematical models \cite{16} fully exploiting the oscillatory instability possibilities of a saddle-node pair of fixed points. The design procedure for $m=1$ is briefly described in the next subsection to illustrate the three facets of the problem: the possession of fixed points, the occurrence of oscillatory instabilities, and the saddle approach to homoclinicity. The task of designing $m>1$ systems fulfilling to some extent the various conditions together might presumably be extremely difficult for a researcher but, perhaps, not so for nature. For instance, we find reason to suspect the occurrence of a high-degree of instability behavior in two relevant problems: the onset of turbulence in moving fluids and the oscillatory activity of living brains. Different aspects of such a possibility for the two cases are considered in the Appendixes B and C, respectively. In fact, phenomena involving a relatively large number of interrelated oscillatory processes with different time scales are ubiquitous. Typical examples may be found in the Earth's climate \cite{57,58}, population dynamics \cite{59}, biological rhythms \cite{60,61} and, although not so conclusive, in economic data and social activities. And, for instance, while the meteorologic and climatic variabilities are currently related to the irregularities of a supposed chaotic evolution and the forecasting limits to its sensitivity to initial conditions, the generalized Landau scenario allows for a more natural interpretation based on the intermodulatory combination of a large number of physical effects with different time constants, whose unpredictability can arise from the lack of a proper description of the involved effects, without necessity of invoking any delicate sensitivity of the system evolution.  

\subsection{Vector fields with a one-directional nonlinear part}

System (1) with \textit{m} = 1 may be usually transformed \footnote{If rank of ($b_{1}$, $Ab_{1}$, $A^{2}b_{1}$, . . . ., $A^{N-1}b_{1}$) is equal to $N$ \cite{62}.} to a canonical form based on the companion matrix as follows
\begin{subequations} \label{equations 2}
\begin{align}
\frac{dz_{1}}{dt}&= -\sum^{N}_{j=1}c_{j}z_{j}+f_{1}\left(z_{1},\,z_{2}\,...,\,z_{N},\,\mu\right), \label{2a}\\
\frac{dz_{j}}{dt}&= z_{j-1}\,,\:j=2,\,3,\,...,\,N, \label{2b}
\end{align}
\end{subequations}

\noindent
whose equilibria appear located on the $z_{N}$ axis. In the absence of any Hopf bifurcation, the one-dimensional array will consist of an alternate sequence of fixed points differing by one in their unstable manifold dimensions and, in particular, we are interested in sequences of $S_{0}$ and $S_{1}$ points, i.e., we want stability outside the line of saddle-node connections to guarantee the generic presence of one attractor before and after the Hopf bifurcations. In this case, the stable manifolds of a $S_{0}-S_{1}$ pair can sustain up to $(N-1)$ different bifurcations.

In order to achieve systems able to sustain the full instability behavior of their fixed points, we consider nonlinear functions of a single variable in the form
\begin{equation}\label{equation 3}
f_{1}\left(z_{1},\,z_{2},\,..,\,z_{N},\,\mu\right)=\mu_{C}\,g\left(\psi,\,\mu\right),
\end{equation}
with
\begin{equation}\label{equation 4}
\psi=\sum^{N}_{j=1}d_{j}z_{j},
\end{equation}

\noindent
and where $\mu_{C}$ will be taken as (a very convenient) control parameter. This kind of function allows us to divide the design of the system in two separate problems: the existence of fixed points and the occurrence of Hopf bifurcations on these points. In effect, although it may appear strange, the linear stability analysis of the fixed points of the families of systems in the form (2)-(4) can be implicitly done without specifying the nonlinear function and, therefore, without knowing the actual fixed points. This is because the influence of the nonlinear function on the Jacobian matrix of a given fixed point is fully described by the corresponding value of the auxiliary parameter
\begin{equation}
p=\mu_{C}\,\frac{\partial{g}}{\partial\psi}.
\end{equation}

\noindent
In addition, the $p$ value identifies the kind of fixed point since it is equal to 1 for the nonhyperbolic fixed point of any zero eigenvalue bifurcation and becomes lower (higher) than 1 for the node (saddle) point emerging from the bifurcation. Let us here briefly recall the two steps of the design procedure \cite{16}. Firstly, after choosing the dimension $N$, the stability analysis of a generic fixed point as a function of $p$ is used to determine the linear part of the system, i.e., the $c_{j}$ and $d_{j}$ coefficients, in order to assure hypothetical fixed points that i) will be of the types $S_{0}$ and $S_{1}$ before the occurrence of any Hopf bifurcation, and ii) will be able to exhaust their stable dimensions through successive Hopf bifurcations at increasing values of the control parameter. This design step is done by (properly) choosing the values for the oscillation frequency and parameter $p$ of the $N-1$ bifurcations of a generic saddle-node pair of fixed points. The second step is to choose the nonlinear function $g(\psi)$ in order to have the wanted fixed points with the selected $p$ values for reasonable $\mu_{C}$ values. In fact, the actual expression of $g(\psi)$ is not so relevant, provided it should describe some sort of hump with positive and negative slopes allowing for the existence of more than one fixed point. The reported numerical results correspond to a periodic nonlinear function, related to the Airy interferometric function of our physical devices \cite{17}, that provide for a multiplicity of saddle-node pairs upon which the oscillatory scenario investigation becomes much facilitated.

With this procedure we are able to obtain $N$-dimensional systems that possess saddle-node pairs of fixed points experiencing up to $N-1$ Hopf bifurcations and that usually exhibit all of these oscillation modes intermittently combined on the time evolution of the attractors emerged from the node points. The process is parametrically well controlled by the scaling factor $\mu_{C}$. For $\mu_{C}$=0, the single fixed point of the linear system is stable (provided a proper design has been done) and, with increasing $\mu_{C}$, it is accompanied by one or more pairs through saddle-node (perhaps pitchfork) bifurcations. In the process the various fixed points experience successive Hopf bifurcations up to exhaust their stable manifolds. Efficient mixing happens automatically for the oscillation modes of both the node and saddle points and, in particular, this occurs because, independently of $N$, the attractor emerged from a node usually grows with $\mu_{C}$ towards a neighboring saddle point and the saddle cycles emerged from it. Thus, the third element required for efficient mode mixing, i.e., the approach to homoclinicity, seems to work automatically in the case of systems (2)-(4) designed to fulfill the possession of fixed points and the occurrence of Hopf bifurcations.

The advantageous use of nonlinear functions of a single variable in the form of Eqs. (3) and (4) does not represent any loss of generality except for the fact that the different $S_{0}$ points of the array experience identical Hopf bifurcations in the linear regime, and the same happens with the $S_{1}$ points. In other words, each fixed point experiences a succession of bifurcations whose frequencies and two-dimensional center eigenspaces are the same for all nodes, on the one side, and for all saddles, on the other side. This means, for instance, that in the full instability behavior associated with a saddle-node-saddle trio, the oscillation modes of the two saddles will appear with equal frequencies and a similar look but at different locations on both the attractor and time evolution waveform. There is also some restriction in the possible values for the oscillation frequencies, in the sense that the system design works better for clearly different values that, in addition, have been properly ordered in relation to the occurrence of the Hopf bifurcations as a function of the control parameter.\footnote{Under improper choice, one may obtain divergent systems with unstable fixed points before any Hopf bifurcation or find circumstances with no compatible systems in the form of Eqs. (2)-(4).} With this caution in mind, the design procedure has no limit for $N$, in the sense that the $N-1$ oscillation modes appear with full amplitude on the time evolution of the attractor independently of $N$. To facilitate the verification of the reported numerical results we give in \footnote{Fig. 2: $c_{q}$= 50, 440, 480, 360 and $d_{q}$= -17, 66, -200, 360. Fig. 3: $c_{q}$= 250, 11080, 104600, 42300, 5680, 13 and $d_{q}$= -36, 660, -14190, 2910, -940, 13. Fig. 7: $c_{q}$= (0.001, 0.38, 39, 550, 4000, 2000, 330, 2)10$^{6}$ and $d_{q}$= (-0.00014, 0.038, -5.9, 68, -570, 303, -55, 2)10$^{6}$. Fig. 9: $c_{q}$= 100, 14400, 100000, 526000, 34300, 695 and $d_{q}$= -20, 2060, -19400, 89000, -8540, 695.} slightly rounded values of the $c_{j}$ and $d_{j}$ coefficients derived for the set of frequencies and $p$ values indicated in the figure captions. The oscillatory behavior exhibited by any of these systems changes slightly when the $c_{j}$ and $d_{j}$ coefficients are modified or when different $g(\psi)$ are employed. This points out the robustness of this kind of behavior and its noncritical localization in the space of the dynamical systems.

\begin{figure*}[t]
\centering
\includegraphics[width=0.8\linewidth]{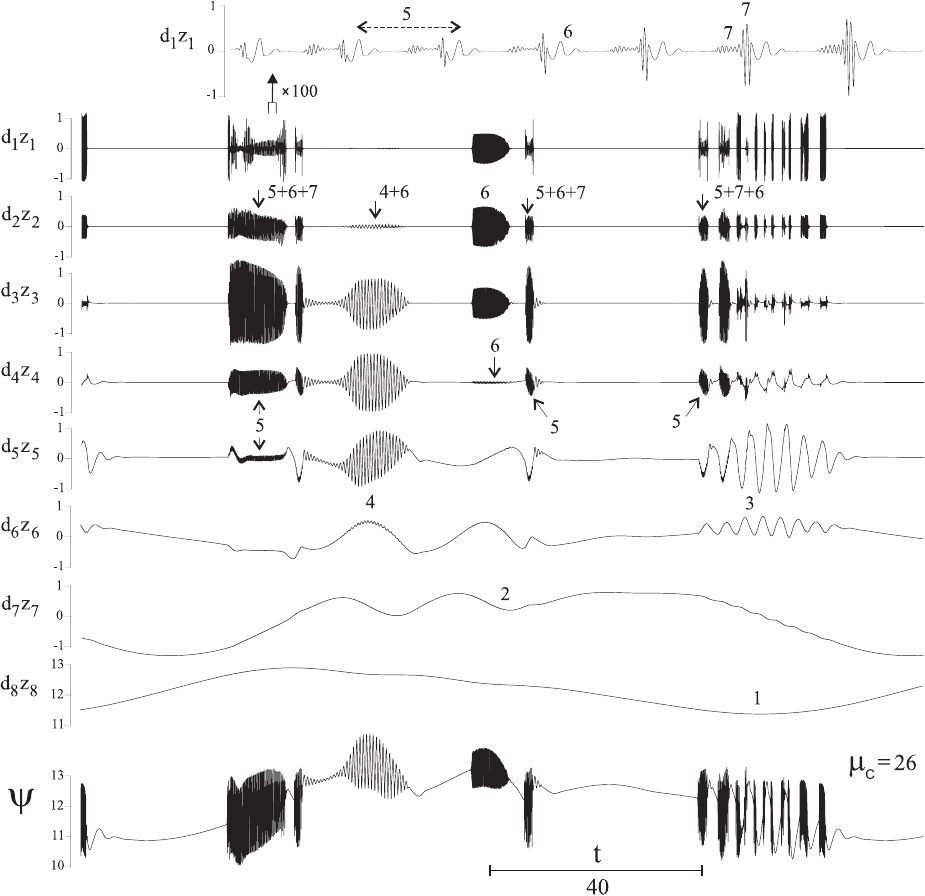}
\caption{The same as in Fig.~\ref{Fig6} but for $\mu_{C}$=26 and including the time evolution of the eight variables of the system in the canonical form of Eqs. (2)-(4). The coefficients $d_{q}$ include an alternatively opposite sign as a function of $q$ that, in addition to the scale, should be taken into account when trying to appreciate the time differentiation relation among successive variables.}\label{Fig7}
\end{figure*}

Numerical and experimental results concerning nonlinear mode mixing have been previously reported \cite{16,17} and here we just illustrate the main features of the oscillatory scenario and present some unreported complementary views like the phase portraits of complex orbits (Fig.~\ref{Fig3}) and the time evolution of the different variables of the system (Fig.~\ref{Fig7}). For this purpose we have chosen systems exhibiting clearly different oscillation frequencies so that a clear visualization of the orbit structure in both the time evolution and phase space is easily achieved. The time evolutions shown in Figs. 7 and 8 correspond to an $N=8$ system. The canonical form of Eqs.(2), with $z_{j}=z^{(N-j)}_{N}$, $j=1,..,N$, where the superscript denotes the order of time differentiation, explicitly illustrates how one of the variables as a function of time contains the full information of the rest of them and this is particularly impressive when considering the really smooth evolution of $z_{N}$ in Fig.~\ref{Fig7}.
\footnote{Notice, however, that the dynamical interrelations work in the opposite direction, i.e., $z_{j-1}=z^{(1)}_{j}$ in Eq. (2b) expresses that $z_{j-1}$ determines the time rate variation of $z_{j}$, not that $z_{j}$ determines $z_{j-1}$ through its time derivative. This is particularly relevant to understand how the noise effects propagate through the variables.} The simple differentiation relation among variables implies that the relative presence of the oscillation modes enhances in proportion to their frequency when considering variables of successively decreasing subscript $j$ and, in this way, facilitates the discrimination of the different modes. This peculiar behavior would become deeply hidden in a system transformation generically providing new variables like arbitrary combinations of the canonical ones. In particular, the variable $\psi$ defined by Eq. (4) has two relevant features making it very convenient as observable: it contains equilibrated contributions of the various modes and is sensitive to the relative position of the fixed points. These features provide the clear distinction between the two basic kinds of mode mixing depicted in Figs. 1 and 2.  

Such different frequency values as those in the reported examples point clearly out how the slowly-varying variables modulate the faster activity of others and how the succession of intermittent bursts could be associated with bifurcations of certain undefined subsystems under the modulating control of the whole system. Notice in particular the long time intervals during which the quickly-varying variables remain practically at the zero value. On the other hand, the clear distinction of frequencies seems responsible for the apparent periodicity easily exhibited by these systems. The basic recurrence of such complex time evolutions, as well as the similarity levels of transitional repetitions, make obvious the strict organization under which the system is evolving and it is worth realizing that the source of order lies just in the structure of causal interrelations governing the system dynamics. By looking in particular at the system of Eqs. (2)-(4), it is easy to appreciate the feedback circuits among variables and their time variation rates, as well as the exclusive nonlinear influence of $g(\psi)$, while competition is implicitly contained in the values of the $d_{q}$ coefficients and more specifically in their alternatively opposite signs, as derived from the design procedure. It is worth noting that all these ingredients have been, relatively easily, implemented in experimental devices for $N$ values up to 6.
\footnote{The so-called BOITAL devices consist of a light beam of constant power illuminating an $N$-layer stack of transparent materials with alternatively opposite thermo-optic effects placed in between two flat mirrors, the input one of which is partially absorbing. Feedback occurs through heat diffusion from the absorbing mirror toward the layers, consequent temperature effects on the cavity optical path and consequent light interference changes upon the heat source; nonlinearity arises through the interferences making the light intensity in the absorbing mirror nonlinearly dependent on the total optical path through the corresponding Airy function; and competition takes place among the opposite thermo-optic effects of the various layers, with the corresponding characteristic times related to the different distances to the localized heat source. The incident light power acts as a scale factor on the nonlinearity strength and provides us with a really useful control parameter (like $\mu_{C}$ in Eq. (3)). The reflected power is affected by the cavity optical path through light interferences and then its time evolution manifests what is happening within the device \cite{17,64,65}.} 

\begin{figure*}[t]
\centering
\includegraphics[width=0.8\linewidth]{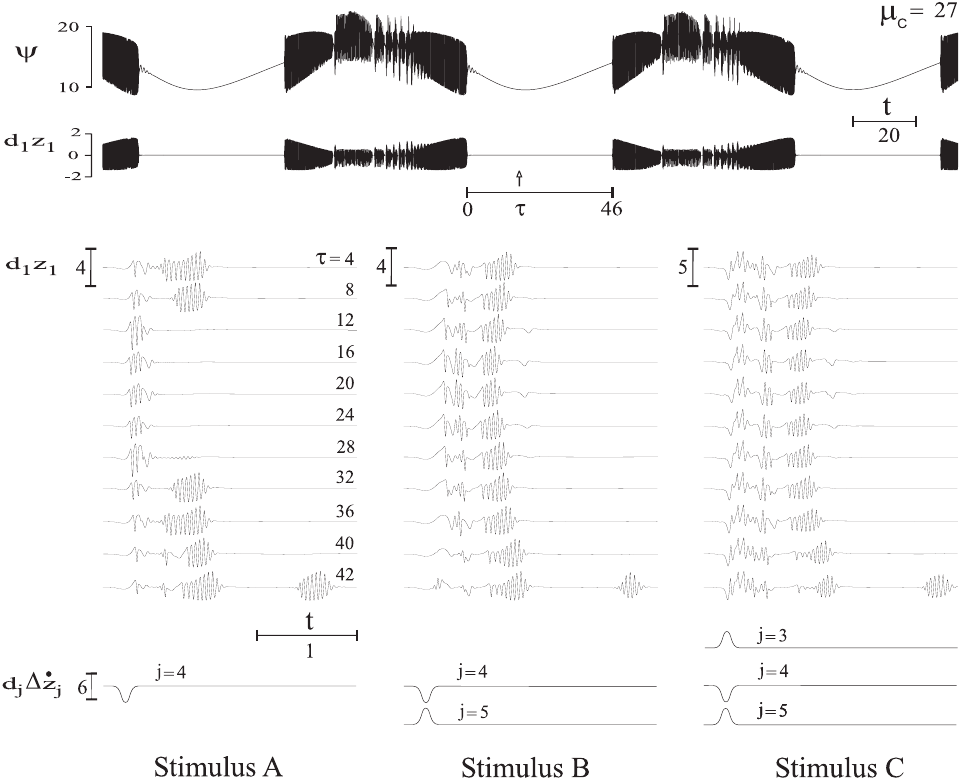}
\caption{Transient behavior of a $\left(N=6, m=1\right)$ system designed by imposing Hopf bifurcations of time periods equal to 100, 10, 1, 0.2, and 0.04, at respective \textit{p} values equal to -4, 6, -5.2, 6.5, and -5, and by choosing $c_{1}=100$. The nonlinear function is as in Fig.~\ref{Fig2}. At the top, the asymptotic time evolution of the attractor for both $\psi$ and $z_{1}$. Each column presents a sequence of transients induced by a given stimulus at different moments of the calm interval, as characterized by the value of $\tau$. The stimuli are Gaussian pulses applied through an additive term on one or more of the dynamical equations, e.g., stimulus B contains a negative pulse on $dz_{4}/dt$ and a positive pulse on $dz_{5}/dt$.}\label{Fig8}
\end{figure*}

At this point, we find instructive to imagine the thoughts of one of those observers who are searching for extraterrestrial life evidences while hypothetically receiving one of such time signals from some remote source, especially when, after a long interval of virtual inactivity, the complex sequence of oscillations repeats just as before, and again and again. In fact, when analyzing any kind of recurrent complex behavior, the observer would probably need explanatory reasons concerning both how the system can determine the complex sequences of its time evolution and how it can indefinitely sustain its dynamical workings without disaggregating, and the instructive conclusion should be to realize an answer to such questions in the general context of the dynamical systems. After this, the observer could attribute a given degree of structural sophistication to the system and consequently raise questions concerning how its assemblage could have occurred, for which, however, there is no immediate answer within nonlinear dynamics.

\subsection{Transient activity around the attractor}

In the absence of external perturbations, a system possessing an attractor will evolve by following the intermittent sequences of oscillatory modes imprinted on the necessarily recurrent time evolution of the attracting state. Nevertheless, the nonlinear mode mixing does not restrict just to the attractor but affects extended phase space regions within the basin of attraction, as determined by the intertwinement of invariant manifolds of the several saddle limit sets. The flow of these regions corresponds to transient trajectories that can be selectively induced by properly and momentarily perturbing the system state and each one of them describes a peculiar oscillatory pattern during its return to the attractor. Without pretending a detailed analysis of the transient repertoire and its relation to the perturbation map, which may be really cumbersome in the $N$-dimensional phase space, we simply report here an example displaying some features particularly useful for the dynamic brain analysis of Appendix C.

In the case of Fig.~\ref{Fig8} the time evolution of the attractor shows the rich oscillatory activity clearly accumulated in between long intervals of calm, and this will allow us to use such a kind of oscillatory pattern for illustrating the intrinsic wake-sleep cycle tentatively assigned to the dynamic brain, in which calm will correspond to waking and complex oscillations to sleeping. We will then consider the action of sensory input during waking as introducing perturbation displacements against which the intrinsic dynamics tends to recover the asymptotic state of the attractor, and we will attribute to such a transient activity the basis through which the waking dynamic brain should operate. Along this line, the example of Fig.~\ref{Fig8} illustrates how the recovering transients are for a variety of stimuli applied on the system when it is found in different moments of the calm interval, although in the simple case of a system with $N=6$ and $m=1$. Notice to what extent the stimuli induce characteristic transient responses according to peculiar features of their perturbation map, and to what extent such characteristic features are independent of the moment when the system is perturbed within the calm region. As discussed in Appendix C, the achievement of these properties would provide a well defined association between input perturbation maps and transient oscillatory complexes, upon which the conjectured dynamic brain could sustain its identification and processing capabilities. Notice also that, while the figure reports the transient signals on one of the variables only, the actual trajectories develop in the $N$-dimensional phase space by specifically affecting the different variables in peculiar ways and this implies a very rich repertoire of transient behaviors. Also relevant should be the analysis of how the transient perturbations propagate among variables of different timescale sensitivity \cite{ROS} and how the induced transients of different frequencies attenuate ones with respect to others.

The system of Fig.~\ref{Fig8} corresponds to a well developed Landau scenario with all its potential frequencies already present in the attractor but, in cases of not so developed scenarios, the transients may exhibit oscillatory modes absent in the time dynamics of the attractor. This reflects the general fact that, when varying a control parameter, the oscillation modes emerge in the phase space around the fixed points having done or being near to do the corresponding Hopf bifurcations and posteriorly extend their influence toward the attractor by affecting the intermediate phase space regions. On the other hand, it is worth stressing that, leaving apart the possible coexistence of attractors, the attainable dynamics of the system corresponds to the full basin of attraction and that there is an intrinsic relation between the asymptotic trajectory of the attractor and the transient flow of the basin, in the sense that any parametric variation of the system modifies the two aspects of the flow accordingly. All these details will be relevant when considering the proposed mechanism of learning for the dynamic brain.

In short, a dynamical system having developed a generalized Landau scenario to some extent is a sort of excitable medium with a varied repertoire of transient behaviors, in addition to the recurrent evolution of the attractor, and we suspect an extraordinary development of such possibilities with increasing the multi-directionality $m$ of the nonlinear part of the vector field. 
 
\section{DYNAMICAL OSCILLATIONS IN A DETERMINISTIC WORLD}

When trying to explain the time evolving features of the observable world through the deterministic paradigm of nonlinear dynamics, the question of what things the dynamical systems can do in order to produce complex behaviors becomes really relevant. Our answer to this question is that, in practice, they can do just one thing: to oscillate. Along this line, the paper tries to establish a generic phase space scenario for the optimum development of the oscillatory possibilities of (dissipative and autonomous) dynamical systems and arrives to what we call the {\em generalized Landau scenario}. In this scenario the system can have robust, and not necessarily high-dimensional, attractors sustaining the nonlinear mixing of a number of characteristic oscillation modes much greater than half the number of degrees of freedom and it has the peculiar feature of expansive growing scalability.\footnote{The presence of an attractor makes the scenario features more comprehensible, but analogous oscillatory mixing scenarios lacking any attractor are also possible.} The name comes from the fact that we interpret the scenario as the way nonlinear dynamics has for developing the physical intuition expressed by Landau through its phenomenological theory tentatively explaining the onset of turbulence like a combination of oscillations \cite{13}. The mechanism described by Landau is essentially linear (except for the stabilization of the oscillations) and the generalization at the nonlinear level is threefold: i) the number of times $N$ degrees of freedom can sustain $N/2$ oscillation modes through different fixed points, ii) the capability to combine (often intermittently) the modes emerged from a (properly connected) set of fixed points into the time evolution of a given attracting limit set, and iii) the extremely rich repertoire of transient oscillatory patterns contained in the basin of attraction in addition to the recurrent one imprinted on the attractor. It is worth stressing that the scenario expresses nothing but the manner in which the circuits of causal influences among properties of both the system and its environment determine the subsequent time evolution of the system properties for different sets of initial values. Notice also that the complexity and diversity of oscillatory patterns arise from the properly organized presence of degrees of freedom, feedback, competition and nonlinearity in such circuits.

\subsection{Emergence and building of complexity}

We consider that the emergence of oscillation modes in the phase space provides the bricks with which the building of complex dynamical behavior may occur and that the generalized Landau scenario provides the frame through which this building can develop. Each oscillation mode represents a complexity step in the system behavior simply because its emergence is a two-dimensional event \footnote{That can, however, engage an arbitrary number of variables in accordance with the phase space orientation of the oscillatory plane.} through which two degrees of freedom become firstly correlated one another in a certain phase space region (by sustaining a spiraling flow in a neighborhood of a dynamical equilibrium) and then, just thanks to such a correlation, perform the Hopf bifurcation like a codimension-one process. \footnote{Notice that no bifurcations of codimension-one and dimension higher than one are known other than those associated with two-dimensional oscillatory instabilities.} That is, the crucial moment in this step occurs when two degrees of freedom become (locally in the phase space and linearly) correlated through the proper occurrence of competing feedback in the causal circuits, i.e., in the case of a mathematical system, when two real eigenvalues of an equilibrium become equal and then a complex conjugated pair. \footnote{The complexification of two eigenvalues is not a reason of anything but a mathematical manifestation of the oscillatory correlation between two degrees of freedom.} The step culminates after the oscillatory instability, when the causal interrelations become (nonlinearly) able to self-sustain the recurrent time evolution of the oscillation mode on the periodic orbit and to extent its influences along the unstable manifold branches in the case of a saddle. 

The nonlinear mixing of the various oscillation modes expresses dynamical correlation among the corresponding degrees of freedom and, while the modes emerged from the same fixed point deal with additional degrees of freedom up to exhaust the value $N$, the modes from different fixed points embody the multiplicity of ways through which the same degrees of freedom can (intermittently) participate in the oscillatory activity of the system. Such a multiplicity of ways arises from the nonlinearities allowing the coexistence of pairs of saddle-node connected fixed points, which are also related to codimension-one events like the single zero eigenvalue bifurcations. Thus, the generalized Landau scenario is based on sequential chains of codimension-one events and this makes its development feasible in practice. It is really worth remarking that the mode mixing processes of the oscillatory scenario cannot develop arbitrarily but under the phase space topological constraints associated with the intertwinement of invariant manifolds, as generically discussed in Appendix A. The consequences of such constraints are twofold: a limitation of available dynamical behaviors and an implicit repertoire of phase-space pathways for the processes of complexity emergence.

We exclude chaos as an effective way for complexity accumulation because we consider the essence of its complex features (i.e., the close coexistence of an indefinite number of unstable periodic orbits and the strange properties of the attractor) excessively subtle for such a purpose, and, even accepting the potential relevance of the sensitivity to initial conditions and the irregularity of chaotic evolutions for natural systems, we don't find any reason to consider them basic mechanisms for building up additional complexity into the dynamical behavior. 

Within the context of deterministic systems, we consider the main dynamical mechanisms underlying the spatio-temporal phenomena in extended systems as included in the possibilities of the generalized Landau scenario. The infinite spatial resolution, so powerful for describing spatially-extended systems through partial differential equations, seems unnecessary in order to pick up the causal interrelations effectively sustaining the dynamical behavior of natural systems, and we assume that proper reduction procedures must then provide behaviorally equivalent systems of ordinary differential equations. Notice that, in addition to the finite number of effective variables, this means also that the dynamical effects associated with spatial propagation and transport processes can be subsumed in the structure of causal dependences and parameter values of the properly reduced system, without explicit consideration of time-delayed influences.

In conclusion, we regard the development of the generalized Landau scenario as the genuine way toward complex behaviors in a hypothetical world whose behavior is supposedly described by nonlinear dynamics, and this convincement has impelled us to consider its potential implication in the turbulence of fluids and the oscillatory activity of brains, as discussed in the appendixes, with a twofold aim: to achieve plausibility support for the scenario and to attempt a plausible approach toward an explanation of two apparently insolvable problems. We don't claim any explanatory findings, because the analyses are based on assumptions pending corroboration, but the achievement of enough plausibility to justify a deeper scrutiny of the overall framework.

\subsection{Structural evolution of systems}

In fact, to keep such plausibility up, we need a coherent answer to the question of how the generation of systems sustaining so complex oscillatory scenarios could have occurred in a supposedly deterministic world. This brings us to more general questions like what the evolution of systems is and why and how it happens. We are referring to the observable fact that the time evolving features of the real world include the occurrence of qualitative transformations in the properties (of things) defining causal relations, as well as the emergence of new causal chains. This dynamically structural facet of the time evolution constitutes the essential feature in the functioning of systems exhibiting complex behaviors and includes what can be interpreted as the formation of subsystems and their posterior evolutionary transformation. The differentiation of a subsystem within a whole acquires sense as far as it attains both enough dynamical activity and enough dynamical autonomy with respect to the interacting environment. In our view, the former means just a certain extent of intrinsic oscillatory activity and the latter means that the influencing properties of the environment remain almost fixed much longer than the endogenous characteristic times. At the ideal limit, full dynamical autonomy would mean strictly fixed environmental influencing properties and then the phase space of the subsystem could be considered independent of that of the whole system.\footnote{Autonomy does not imply fixed environmental influences on the system because such influences happen through the interaction with internal properties that can be evolving in time, i.e., the external influences vary under internal control in an autonomous system. Autonomy means also that the possible time-dependent influences of the system on the environment do not feed back to it.}

Notice the duality of dynamical complexity between activity and structure in their relation to the concept of organization. The structural organization resides in the circuits and functional dependences of the causal interrelations among the set of properties effectively sustaining the dynamical activity of a given subsystem. Such an activity is nothing but the time evolution of those properties whose value significantly changes, and the underlying organization manifests in how each one of these evolutions will occur along the time and in how each one of the evolving variables is in relation to the others at every moment. In fact, to be precise, the activity refers to the values of all the properties participating in the dynamical interrelations, including those remaining practically fixed (usually called parameters in nonlinear dynamics). Structural organization does not imply alterations in the laws of the causal (physical) effects sustaining the dynamical interrelations but a properly organized interplay among the workings of such effects from which peculiar behaviors arise in the activity. The issues of the structural evolution of systems lie in i) how the organized activity can influence the underlying structural organization through the transformation of dynamically-relevant properties and of their interrelations, ii) why such an influence is generically directed toward an increase of organizational degree, and iii) how such an organization growing provides inherent persistence to the emerging structures. These issues define the problem of self-organization \cite{19,18}, which cannot be interpreted literally as a system organizing by itself but in the context of interacting subsystems that organize ones to others, typically seen as one system and its interacting environment.\footnote{Supposing definite spatial boundaries among subsystems makes the analysis easier but such a view is not generic enough and, in addition, it depends on the employed definition of system. In fact, in the paper, we leave the meaning of the term system relatively ambiguous to avoid unnecessary additional abstraction.} On the other hand, by introducing some relation between structural organization and the organized accumulation of information, the same issues are part of the problem of learning and memory. In fact, the evolutionary framework we will now present has arisen from the learning framework developed for the dynamic brain in Appendix C.

By considering the world as a deterministic dynamical system, without external influences, and independently of whether it evolves either on a transient or close to an asymptotical state, (and being convinced of the singular role of the generalized Landau scenario for the building of dynamical complexity), we must associate the occurrence of dynamically structural transformations in a region of the real space with the local development of complex oscillatory activity in the course of the time evolution of the whole system. Thus, we interpret the transformations as arising from processes through which the interacting parcels of the whole induce one another to modify their circuits of causal relations so as to achieve higher levels of oscillatory complexity. In the building of the mode mixing scenario, some of the parcels can enhance their dynamical autonomy in the path to becoming subsystems. Similar reasoning would apply to the evolutionary transformation of a relatively autonomous subsystem toward the formation of internal subsystems. Such a kind of interpretation requires complementary ingredients of two types: ones of general quality favoring the formation of local dynamical structures through the global stream of causal influences, i.e., generic reasons giving likelihood to structural development within the strictly deterministic time evolution of the whole system, and others of particular character describing the {\em appropriate circumstances} occurring where and when the structural evolution of systems takes place.

In addition to the elementary steps and prefigured courses of complexity building tentatively associated with the generalized Landau scenario, we perceive reasons for a generic asymmetry of opportunity between the building and deconstruction of mode mixing potentially occurring in a given parcel of reality in interaction with its environment. In other words, we perceive a directional evolutionary arrow arising from reasons making more likely the externally-induced incorporation of a new oscillation mode to the mixing with a previously existing mode than the externally-induced removal of that mode. In this way, we could expect the (natural) occurrence of local accumulations of dynamical complexity in a deterministically evolving world. One of the reasons is inherently associated with the dynamical correlation of two degrees of freedom embodied in each oscillation mode and the consequent asymmetry of requirements for the stimuli actuation on opposite directions, i.e., in simple terms, the mode inhibition would require two-dimensional stimuli in proper (anti) synchrony with the internal oscillation, while the induction of absent modes does not require any synchrony at all. This description corresponds to situations in which the induction or inhibition of the oscillation mode happen directly on the attractor but, in general, the mode emergence or submergence should occur nearby the fixed point experiencing the corresponding Hopf bifurcation. This fixed point usually would be a saddle relatively far from the intrinsic asymptotic trajectory of the transforming subsystem, and then the external perturbation must actuate in a twofold way by properly displacing the subsystem state and by either inducing or inhibiting the oscillatory motion.  We find here strong reasons for the directional arrow because the sequential requirement of proper displacement and two-dimensional inhibitory synchrony looks very unlikely. 

At this point, a more concrete picture of the evolutionary mechanism and of what we are calling {\em appropriate circumstances} for it is needed. For this purpose, under the guide of the dynamic brain analysis of Appendix C, we introduce the rough concept of \emph{ plasticity} by hypothetically associating it with undefined subsystem features that can be altered in a lasting way by the proper momentary action of external stimuli. The {\em appropriate circumstances} for dynamically structural evolution would refer to both the transforming subsystem and interacting environment, and they could be generically described as consisting in: 

\begin{itemize}\setlength{\itemsep}{-4pt} \setlength{\topsep}{0pt}
\item[1)] The presence of plasticity in the evolving subsystem, properly working in the threefold sense that, first, the external stimulus does not act directly on the plastic features but through the oscillatory activity of other variable properties of the subsystem; second, there is positive feedback between the plastic changes and the oscillatory activity sustaining them; and, third, the plastic changes remain when the external stimulus vanishes and the given oscillatory activity eventually attenuates. The intermediate oscillatory activity is essential since it would be the oscillation mode involved in the complexity step, while plasticity would sustain the lasting transformation imprinting the oscillatory motion in a certain region of the phase space. Positive feedback is required to promote the mode emergence and, in addition, it could provide for inherent persistence of the complexity step.
\item[2)]	The occurrence of proper stimuli from the environment in order to perturb the system state and induce oscillatory traces of the emergent mode in the corresponding phase space region (of the nearly autonomous system in the absence of stimuli). The adjustment and maintenance within proper ranges of the steady  environmental properties (i.e., those changing slowly with respect to the system dynamics) is also needed in order to preserve the phase space scenario features.
\end{itemize}

All such circumstances should occur in the course of the time evolution of the whole system and, in principle, appropriate circumstances for mode inhibition would also be possible. However, besides the difficult requisites for the inhibitory stimuli previously noted, we find unfeasible the achievement of positive feedback between plasticity and mode inhibition when the oscillation mode is not continuously working, i.e., when it must be excited by the inhibitory stimulus. These reasons for asymmetry of opportunity between building and deconstruction of oscillatory mode mixing could explain the spontaneous occurrence of local accumulations of dynamical activity in a supposedly deterministic world, provided that the set of appropriate circumstances for constructive building would actually occur to a certain extent and this should be a feature intrinsically related to the dynamical possibilities of the actual constituents of that world.

The working of the evolutionary arrow is not in contradiction with the finite duration often observed for the emergent subsystems, since the disappearance does not happen through mode mixing deconstruction by external stimuli (i.e., the opposite of building) but usually through gradual variation of some of the (quasi) steady external influences. An example for this could be the formation and extinction of hurricanes \cite{71}, in which the formation requires something more than the proper steady environmental properties provided by the ocean and atmosphere, while the extinction occurs when such steady influences gradually change due, for instance, to inland penetration. In fact, such a kind of asymmetry illustrates the main distinction between the notions of birth and death, as applied to complex dynamical structures of different nature, in the sense that birth always implies development by cumulative growing while, among its wide variety of manners, death rarely (perhaps never) involves gradual deconstruction of the complex system. Of course, the key question is what accumulates in the developmental growing of a complex system and whether such an accumulation has common features for different systems.

The evolutionary mechanism we are conjecturing here identifies the local emergence of dynamical complexity, in its behavioral and structural facets, with the accumulative nonlinear mixing of oscillatory modes in the (probably transforming) phase space of the evolutionary subsystem and with the underlying development in the physical space of properly organized causal circuits among properties of both the transforming subsystem and its environment, respectively. We sustain it on three complementary ingredients: i) the elementary steps and prefigured courses of complexity building associated with the generalized Landau scenario, as allowed by the generic possibilities of the dynamical systems, ii) the evolutionary arrow propelled by the asymmetry of opportunity between the occurrence of appropriate circumstances for either building or deconstruction of nonlinear mode mixing, and iii) the presumption that the intrinsic evolving features of the world effectively include the local and temporal occurrence of appropriate circumstances for constructive oscillatory building.
 
A theory of the evolutionary mechanism should include a characterization of the appropriate circumstances for oscillatory building and it will be of practical interest only if the wide variety of conceivable particular situations can be subsumed into a single (perhaps a few) generic depiction. This is what we have tried to achieve in our description above, in which, however, the decisive concept of plasticity appears implicitly defined and requires further clarification. The appropriate circumstances include the role of what other evolutionary approaches attribute to chance. We consider the occurrence of appropriate circumstances as causally sustained but independent of the randomness degrees of the underlying sources from which the involved physical processes take place. On the other hand, the observable tendency of accumulative development in systems exhibiting structural evolution suggests that the evolutionary efficiency, i.e., the presumed intrinsic occurrence of appropriate circumstances for oscillatory building, could become enhanced with the evolutionary enrichment of the interactive pathways between a given subsystem and its environment.

The characterization of appropriate circumstances, if successfully done, would represent the useful part of the evolutionary theory because it embodies the exclusive connection to the actual evolutionary system and could enlighten the conditions under which a given emergence process occurs. At this point one is quickly tempted to consider hurricanes, tumors, coronary plaques, and so on, but also quickly realizes how difficult the analysis of a particular case is. Notice that the presumed evolutionary mechanism operates step-by-step, with very elementary steps, and this means that it could be very general, perhaps universal, but also that its relation with the global emergence process yielding a specific subsystem might be very far from obvious, as well as it might be difficult to identify a definite starting moment of the process.

To tentatively sustain such a general role for the hypothetic evolutionary mechanism we would need to establish, on the one hand, the relation of the dynamical activity undoubtedly underlying any complex behavior to the given oscillatory scenario, and this at any scale or level of observation, and, on the other hand, the relation of the emergence process of the dynamic structure responsible for any of such behaviors to the proper occurrence of {\em appropriate circumstances} in both the active subsystem and its environment. We cannot confront here such a widespread view but we find significant that the relatively detailed analyses of Appendixes B and C dealing with turbulent flows and living brains, respectively, connect well with it. In this regard let us remark the following points:

\begin{itemize}\setlength{\itemsep}{-4pt} \setlength{\topsep}{0pt}
\item[a)] The oscillatory conjecture for explaining turbulence acquires a peculiar relevance because the aggregation of matter in the fluid phase seems to have been very common from the very beginning of the world history, and because the fluid aggregation processes would have surely not occurred at the rest state but under significant relative motions that, according to our conjecture, would be enough for developing the oscillatory scenario of turbulence. This suggests a potential role of the oscillatory activity in the processes of matter aggregation, whose working is in general not fully understood and in which intriguing contributions of inner plasticity and environmental effects come always about, as may be illustrated with apparently simply problems like how two drops coalesce \cite{72} or with the ubiquity of catalysis in any kind of combinative reactions.
\item[b)] The development of turbulence with increasing the Reynolds number does not imply any structural evolution since the dynamic interdependences within the flowing fluid do not need to change, i.e., in simple terms, the Navier-Stokes equation remain unchanged with varying the Reynolds number. Thus, it would be the meeting of flowing motion with proper plasticity in the fluid medium which could be potentially relevant for the evolutionary emergence of novel dynamical structures.
\item[c)] A brain having the learning capabilities hypothetically assumed for the dynamic brain could be interpreted as a piece of matter where the occurrence of appropriate circumstances for oscillatory building has reached a really high degree and this refers to both the inner properties and the interactive pathways with the environment. This brain learning would imply dynamically structural evolution and it could be considered reminiscent of the ontogenetic brain development, as well as it could be related with the phylogenetic evolution yielding larger and powerful brains. This provides feasibility to the arbitrarily assumed plasticity learning rule that, as noted in Appendix C, would represent the most demanding merit of the hypothetic dynamic brain.
\item[d)] By viewing structural evolution as concerning the emergence of novelty in deterministically evolving systems, its presumed profusion in the brain could be a way for the achievement of superior functions like the creative and decisory facets of cognition.
\end{itemize}

At this point we consider our tentative theoretical framework coherent enough to make plausible the existence (by evolutionary development) of dynamical systems with high oscillatory capabilities in a supposedly deterministic world. We say this without having being able to illustrate the evolutionary mechanism with any particular case and, in this respect, a comment concerning the micro- side of the matter aggregation problem is worthwhile because it looks in principle as the most appropriate for developing a detailed analysis. We choose atoms and molecules as our more comprehensible microscopic level and find reason to consider that any one of them entails a higher degree of dynamical activity than its separate constituents, that such an activity apparently looks of oscillatory nature, and that such an activity is just which provides the system with building possibilities of additional complexity. On the other hand, when considering the electron capture by an ion to form a neutral atom or the meeting of two atoms to form a molecule, it is known from physical conservation principles that the final stability of any of such processes requires the participation of a third element in interaction with the gluing constituents, in what may be understood as a necessary influence of the environment, i.e., the origin (or receiver) of the so-called spontaneous emission if there is photoemission or the third body itself if there is a three-body encounter. Concerning the role of the third element in relation to our framework, we would need to realize how it participates in inducing new oscillatory activity, how through such actual oscillations it influences certain (plastic) features of the aggregating system up to lastingly consolidate its full phase space scenario, and how such a consolidation involves positive feedback. We would need to realize also that the fragmentation of the aggregated piece, which is possible of course and which implies the participation of a third body also, has in general less likelihood to occur than the aggregation and that such asymmetry arises from reasons related to the third-body roles. 

All of these items are of difficult analysis, as commonly happens in matter aggregation problems, but at the micro- side we meet an additional difficulty arising from quantum mechanics because this theory does not provide us with a dynamical systems perspective. The issue has no direct relation to uncertainties and probabilistic interpretations since the evolution equations with which the theory associates the time dynamics of a quantum object are deterministic differential equations, i.e., the Schr$\ddot{o}$dinger equation typically. The problem is to interpret in light of nonlinear dynamics what such equations describe in agreement with the experiment. From the dynamical point of view, the most significant atomic feature is the wide set of characteristic frequencies manifested by each kind of atom through its resonant interaction with electromagnetic radiation ranging from radiofrequency to X rays. The theory explains the values of such frequencies all at once by associating them with the energy differences between pairs of energy eigenstates and gives to each one of them a physical meaning in the form of mechanical oscillations at the given frequency when the atomic state contains a superposition of the corresponding pairs of eigenstates. The theory establishes also that an arbitrary state can be described as a linear superposition of the energy eigenstates so that the generic autonomous evolution will correspond to a linear combination of oscillations at the different characteristic frequencies. It is worth noting, however, that, although the frequency/energy relation arises directly from the wave equation, the characteristic oscillations do not appear in the complex wavefunction but in its quadratically-related probability density, in the form of spatio-temporal features that vanish under space averaging, and, for certain pairs of superposed eigenstates, they manifest themselves in the temporal evolution of the (space averaged) expectation value of observables like the electron distance to the nucleus or the electric dipole. Thus, the characteristic oscillations arise as a wavefunction interference effect mediated by the observable operator and this is pointed also out by the lack of oscillations when the atom is just on one of the energy eigenstates, by how the oscillation amplitude enhances with the superposition degree up to the optimum value just at fifty-fifty, and by how certain pairs of eigenstates sustain observable oscillations while others not, as expressed by the electric dipole selection rules. The problem is that, even in the ideal (in fact, unreal) case of conservative atoms, the intricacy of the situation makes us unable to define a proper phase space (and the corresponding dynamical system) where the oscillatory possibilities of one of these extraordinary electromechanical devices could be interpreted. We are unable to assess the dimension or number of effective degrees of freedom and, underlying the problem, there is the linear/nonlinear issue, in the sense that certain representations could introduce nonlinearity into the wanted dynamical equations in contraposition to the linear wavefunction relationship of the wave equation. Furthermore, additional handicaps arise from the difficulties quantum mechanics has in including the electromagnetic dissipation of the atom into the dynamical differential equations and from the fact that the quantum electrodynamics approach moves even farther from the dynamical systems view. Particularly significant is the instability by spontaneous emission of the excited stationary states, necessarily assumed and phenomenologically introduced in quantum mechanics to justify the generic tendency of the atomic electron to be on the ground state, and which strongly reminds us of the stability issues in nonlinear dynamics. In brief, while classical mechanics has been one of the pillars of nonlinear dynamics, the dynamical systems perspective of quantum mechanics is essentially lacking and it seems hard to be attained because lies in the midst of the interpretation problem of the theory.

\subsection{Pending questions, mainly philosophical}

We are aware that our analysis has become immersed within old philosophical problems, with ontological and epistemological aspects, and concerning queries on spatial scales or levels of description, emergent or reductive connections among such levels, uniqueness and universality of time, degrees of causality and determinism at different levels, meaning and origin of irreversibility, proper definition and use of concepts and enough coherence of reasoning in the analysis, and so on. We lack here conditions to tackle any of such queries in the proper philosophical context and, referring the reader to related literature \cite{18,74,75,76,77}, we simply state that, in our view, our considerations about the evolution of systems entail the superposition of four assumptions. The following three: (1) that the observable changes in the evolving reality arise deterministically from causal influences among intrinsic properties of the existing things; (2) that the causal interrelations lead to so well-defined consequences that, independently of the place and scale of observation, the directional stream of consequent changes introduces the common course of what we call time; and (3) that, at a given scale, the observable behavior of a parcel of the whole can be effectively, although coarsely and perhaps temporarily, described through a finite set of quantifiable magnitudes characterizing the participating properties of both the considered system and interacting environment in the underlying stream of causal dynamics, and that such a participation can be lawfully expressed by means of influences on the time evolving rate of certain magnitudes as a function of the values of other ones at the given time. Thus, in short and considering autonomous subsystems for concreteness, with (1) to (3) (except for "although coarsely and perhaps temporarily" in (3), which has to do with the fourth assumption) we are assuming that the possible dynamical behaviors to be observed in a natural autonomous system correspond to the mathematical possibilities of a generic differential equation like Eq.(1), where $N$ would stand for the effective number of degrees of freedom and where autonomy would mean that the magnitudes describing environmental properties remain fixed in time. 

By extension, we can apply the same reasoning to any non-autonomous subsystem of a wider autonomous system and this opens a relevant point in the analysis by questioning whether the variation of the external influencing properties could represent structural evolution of the subsystem or could not. The question can be enlarged to include the consequences of external stimuli on the values of subsystem internal properties that remained constant up to that moment. According to our distinction between structure and activity, and if we refer to mathematically defined systems, the answer must be negative for both cases because the values of the dynamical properties do not participate in the organizational structure but in the activity of either the subsystem or the environment. \footnote{The common idea that externally-induced variation of parameters would provide for structural transformation in non-autonomous subsystems arises from the practice of using parameter values to change the investigated system and from the achievement of qualitatively different behaviors in this way. It corresponds to the transformation of a system by a researcher under the guide of its observable behavior, in which the complexity of one of the subsystems makes the situation very far from the basic evolutionary mechanism we are searching for, and in which the structural transformations could be attributed to the researcher brain exclusively.} The euphemistic term {\em mathematically defined system} refers to (physical) systems in which both the properties and the mechanisms actually determining the effective dynamical interrelations remain unaltered in the time evolution, so that their behavior can be described through a given set of equations, i.e., it excludes the working of plasticity as a way for the incorporation of additional physical effects into the dynamical circuits of interrelations and for the transformation of the effectively involved properties. Thus, in the absence of plasticity, the externally-induced occurrence of a bifurcation, with the corresponding qualitative change of behavior, does not represent structural evolution of the bifurcating subsystem, independently of its dynamical dimension and of the external inductor features. This refers in particular to the formation of spatial patterns in extended systems, which impressively denotes the emergence of organized high-dimensional activity and which one could be tempted to associate with structural evolution into the system, but the (mathematically defined) system has been externally changed in the value of the parameter inducing the bifurcation only, without altering its structural organization, and a simple externally-induced reversion on that parameter should recover the original system exactly.

Notice, however, that the mathematically defined systems exhibit irreversibility of behavior under externally-induced parameter variations and we need to clarify its reasons and consequences, as well as its potential connection to structural evolution. The core of irreversibility lies in the existence of one or more attractors with the corresponding basins of attraction, which change simultaneously with the varying parameter because they are abstract, not actual, features of the system, and in the fact that, instead, the system state needs some time to actually transform itself when trying to follow the changing attractor of the basin where it is found. Thus, while the existing attractors show reversibility, the trajectories of the system state in the back and forth parameter sweeping are generically different for three kinds of reasons: first, because the system state never reaches the changing asymptotic state; second, because the parameter variation affects the several variables in temporally different ways that do not change in sweeping back, and third, because in certain cases there is hysteresis. The first two reasons affect any parameter sweeping, even when the transforming system is able to return to the same but changing attractor and reversibility in the system state is then approachable by properly adjusting the size and timing of the discrete sweeping steps. The third reason is no so general because requires nonlinearities allowing coexisting attractors. Hysteresis occurs when the attractor where the system is disappears because the corresponding asymptotic solution meets with another one and either vanishes or subcritically destabilizes, and then the system state is suddenly found in the basin of attraction of a distant attractor and evolves toward it. The given distance to the attractor makes such a transition really irreversible but the underlying reason is just the same as before: the unidirectional sense of any transient trajectory toward the attractor, and hysteresis occurs because this attractor does not disappear when the old one is recovered by sweeping the parameter back and remains up to the other side of the cycle. It is worth noting that both irreversible parameter sweeping and hysteresis cycles occur in systems of any dimension, including dimension one, and also that, in any case, the strict reversibility of the attractors assures that, waiting enough for effective asymptotic approach, the system would be again in the same state after proper back and forth parameter sweeping. Thus, the behavioral irreversibility of mathematically defined non-autonomous systems under externally-induced parameter changes, including the memory of hysteresis, has nothing to do with any structural evolutionary arrow tentatively providing for the accumulative emergence of complexity.

\subsubsection{Connecting with thermodynamics}

Thermodynamics concerns the macroscopic description of systems possessing a huge number of degrees of freedom at a smaller scale, and it is able to properly deal with phenomena implying such a microscopic activity without its explicit consideration other than introducing the concept of internal energy and characterizing its exchanges with the environment under the definite circumstances of thermodynamical equilibrium. From the dynamical systems point of view, the usual distinction among thermodynamical equilibrium, near equilibrium and far from equilibrium requires clarification because it is unclear to what extent the distinction arises from the system state or from the system itself. The state of thermodynamical equilibrium requires a spatially uniform temperature and this implies restriction to systems with homogeneous boundary temperature or to isolated systems lacking boundary influences, as well as additional restrictions can arise from mechanical, chemical and compositional uniformity reasons. Thermodynamic equilibrium is usually investigated in nearly linear systems, although it could be in principle compatible with nonlinearities. Linear systems generically posses one and only one asymptotic solution, necessarily a fixed point, and this means that, if stability is assumed for physical reasons, a nearly linear system with homogeneous boundaries must posses one stable thermodynamic equilibrium state. Near equilibrium thermodynamics refers to linear or nearly linear systems with inhomogeneous boundaries, so that they posses a single fixed point, assumed again stable for physical reasons, which corresponds to a certain spatially variable but static temperature distribution. Far from equilibrium thermodynamics refers to clearly nonlinear systems, perhaps with homogeneous but usually inhomogeneous boundaries, in which a multiplicity of asymptotic solutions probably coexist, each one of them within the known repertoire of nonlinear dynamics and with the possibility of several attractors. Among these attractors the thermodynamic equilibrium state is in principle possible in cases lacking inhomogeneous boundaries, although it could have become unstable and even disappeared with increasing the nonlinearity strength.

From the dynamical systems point of view, there is absolutely no reason to distinguish the fixed point of a thermodynamic equilibrium state from those of near equilibrium and far from equilibrium systems, all of them describing a dynamical equilibrium through which an autonomous system in interaction with its environment is maintaining its properties constant in time and the stability features of which may be generically equal. It is then relevant to understand the physical reasons for such a drastic distinction in the building of thermodynamics. Besides the practical advantage of managing single values for the homogeneous system properties, the distinctive reason for the thermodynamic equilibrium arises from the peculiar circumstance that a system in such a state does not sustain macroscopic exchanges with the environment, of anything, and of energy in particular.\footnote{See \cite{79}, p. 54. The peculiar static features of thermodynamic equilibrium in relation to the generic dynamical equilibrium are often not clearly stated in textbooks on physical thermodynamics and the consequence is a confusing identification of the particular case with the general one, i.e., a state in which the system properties show constant values is not a proper definition of thermodynamic equilibrium.} This circumstance is manifested when considering ideal infinitely slow processes in which the system transforms under external control by exactly maintaining its state on the transforming fixed point. In the case of thermodynamic equilibrium, the energy exchanges with the environment can be directly associated with the internal energy changes implied in the transformation of the equilibrium state and this is just the way through which the macroscopic thermodynamical description quantitatively connects with the unspecified microscopic activity and it is, in fact, the way around which the edifice of equilibrium thermodynamics is built. On the contrary, this kind of energetic relation becomes indeterminate when the equilibrium state is not a thermodynamic equilibrium, because, in addition to the energy exchanges involved in the ideal transformation of the system in equilibrium state, there are exchanges for maintaining the equilibrium itself at every moment. Such equilibrium inherent exchanges do not reverse when imagining the ideal opposite transformation of the system so that, unlike happens in the thermodynamic equilibrium case, the infinitely slow processes are not reversible and the characterization of the micro-macro connection is obstructed. In a more general view, such a connection does not work neither during the transient states irreversibly evolving toward the attractor, whether a thermodynamic equilibrium or not, which occur either when a system initially on the attractor is undergoing a real transformation or simply when, for some arbitrary reason, the state of an unchanging system is found far from the attractor. Here lies a significant difficulty for extending thermodynamics outside of (thermodynamic) equilibrium.

Notice that, in light of dynamical systems, the irreversible asymptotic tendency toward the attractor corresponds to the evolutionary content of the second law of thermodynamics, in reference to isolated systems possessing a stable thermodynamical equilibrium but which, before becoming isolated, have been externally placed far from such an equilibrium but within its basin of attraction. The same evolutionary reason applies to any autonomous system possessing an attractor, which can be a fixed point, like the equilibrium state of a near equilibrium system (for instance, the linear temperature gradient in a rod with different temperatures at the extremities) or any of the stable equilibrium states of certain far from equilibrium problems (for instance, the B\'{e}nard convection cells in a fluid layer under different top and bottom temperatures), but also a stable periodic orbit or any other kind of attractor in the nonlinear circumstances of far from equilibrium systems. The explanation of the irreversible time evolution of (the state of) these autonomous systems requires, first, the proof of the attractor existence and, second, the establishment of the conditions under which the external and momentary manipulation of the variables of the system leaves it within the basin of attraction of the given attractor, and this necessarily implies particularization of systems. In this sense, a statistical mechanics explanation for the specific case of a gas of moving particles within an elastic box should be a probabilistic proof of such two points, by establishing the required coarse-grained levels of spatial and temporal description of the macroscopic system in order to hidden the statistically relevant fluctuations of microscopic origin within the phase space trajectories and, in particular, within the fixed point of the thermodynamic equilibrium. In other words, a description under which the time evolution initiated from an arbitrary microstate with a given total energy should end in the same macroscopic state of time-constant, spatially-uniform density and temperature. The relative existence of attractor, according to the observation scale in this case, is something inherent to the tentative association of a dynamical system to any piece of reality, in which a proper definition of the dynamically relevant variables is a crucial step (as an example non implying thermal motion, consider the stable fixed points of fluid flows discussed in Appendix B). The attractor relativity for the gas of particles illustrates how the problem of the origin of thermodynamic irreversibility can be identified with that of the origin of attractor at the macroscopic scale whereas each one of the microstates lacks it. In this view, however, we don't find reason for a contradiction with the time reversibility of the fundamental equations of mechanics, nor for connections to any macroscopically specific source of time's arrow.

By recalling that any (mathematically defined) autonomous dynamical system maintains untouched its structural organization during its time evolution, including of course the irreversible pathways towards the attractor, we can realize how the order/disorder scale normally associated with entropy has nothing to do with the structural organization degree of the system. This conventional notion of order is related to the amount of nonuniformity in the spatial distribution of the system physical contents and, while its value varies along the system time evolution, the structure of dynamical interrelations governing that evolution does not. In words, entropy can characterize the state of a dynamical system but not the system itself and there is no way to relate it to the structural organization of the system. This does not mean that the structural evolutionary processes cannot be subjected to restrictions arising from the second law, like to those surely arising from energy reasons and, more in general, from the physical laws governing the working of the dynamical interrelations. On the other hand, however, one could also ask if the applicability domain of the second law is general enough to embrace the structural organization and its evolutionary development. In fact, the question mostly concerning our analysis should be to what extent the second law applies to the supposedly isolated system of the entire universe.

Classical thermodynamics was founded on strict determinism at the macroscopic level but implicitly assuming sources of randomization among the unspecified microscopic degrees of freedom to which the internal energy is attributed. The increasing entropy tendency in isolated systems, as well as the very existence of a stable thermodynamic equilibrium state for such systems, should be associated with the dominance of such a microscopic capability. The same reason sustains the different evolutionary criteria derived from the second law. Statistical mechanics has tried to incorporate equivalent ontological views on its concrete microscopic descriptions, but significant issues concerning ergodicity, rerandomization, and irreversibility remain open \cite{75}, essentially expressing problems of the ways through which a tendency to disorder is tentatively introduced. Thermodynamics was developed by dealing with situations lacking any structural evolution but when, on the way to generalization, the consideration of non-autonomous closed systems suitable for describing ideal reversible processes was extended to the peculiar concept of isolated subsystems, the supposedly isolated entire universe was consequently incorporated into the second law domain with all its features, including structural evolution. The strict universalization of the second law ontologically implies the inexorable dominance of the supposed microscopic randomizing capability as the ultimate reason for the evolution of the world. Under this view, it is difficult to imagine structural evolutionary processes independent of such a microscopic tendency to disorder. The problem refers to both creative and destructive processes and, while the genesis remains pending of explanation, it is today clear that the deterioration (aging) and the cease of activity of structures exhibiting complex behaviors are rarely attributable to any disorder tendency of their constituents. 

Leaving apart our doubts about the proper foundation of the second law universalization and even considering the thermodynamics ontology tentatively compatible with the temporary development of structural organization in a given subsystem, while a larger one including it is on the way to thermodynamic equilibrium, we find unfeasible that this basis can provide foundation for an explanation of such evolutionary development, as it cannot explain what structural organization is and how it works. In summary, we find different kinds of reasons to become skeptical in relation to the possibilities of the thermodynamical approach for explaining the evolutionary traits of the world where we are thinking.

\subsubsection{On the supposedly deterministic world}

Concerning the entire universe, the logical assumption of isolation would imply the absence of boundary sources of inhomogeneity, but the clear presence of nonlinearities means a far from equilibrium system and then the existence of a stable thermodynamic equilibrium state is questionable. To elucidate features of the phase space where the supposedly deterministic world could be describing its actual trajectory, we must consider the known, or at least commonly accepted, features of that trajectory, among which the main detail for us is the continuous occurrence of dynamical organization of matter at all the scales. In our view, a dynamical system describing the trajectory we associate with such a kind of behavior necessarily requires a phase space with complex structures of invariant sets, like those we attribute to the nonlinear mode mixing of the generalized Landau scenario, and, although the existence of attractor should not be sure, in that instance, we would consider very unlikely the possibility of a stable fixed point unaffected by the dynamic interrelations sustaining so complex activity in its basin of attraction. In fact, if an attractor should be there, we would expect it like one of those generically associated with the generalized Landau scenario, in which the asymptotic state would evolve necessarily describing recurrent repetitions. In that case, for discretional reasons, we would prefer to consider our supposedly deterministic and isolated world not in a transient but just on the asymptotic state in order to avoid the insoluble philosophical problem of Before the Beginning in addition to that of After the End.

We are now entering into a crucial but delicate point by asking if the workings of the world involve something more than the physical mechanisms arising from its elementary constituents and their interactions. In light of our evolutionary framework, these additional things could include those defining the organizational possibilities and the development of effective dynamical interrelations at different scales. The relevant question should be if the rules governing these things are independent of those regulating the matter constituents and their interactions, even though the actuation of any dynamical interrelation takes surely place through such physical interactions exclusively. In fact, our evolutionary framework entirely develops within this delicate domain \footnote{Current theories of biological evolution do not trespass on such a domain by hiding the responsibility of novelty emergence under the noisy chance of mutations and the defining sieve of natural selection. At the best of our knowledge, was the Kauffman's proposal \cite{21} the first to penetrate into this domain by attributing definability to the sources of spontaneous order tentatively associated with the dynamical scenarios at the edge of chaos. On the other hand, when formulated as an evolutionary criterion involving the statistical interpretation of entropy, the second law of thermodynamics defines a macroscopic tendency for the future of systems independently of their physical substrate and, usually, this has not been considered a problem by tacitly accepting the microscopic tendency to disorder as arising without requiring concrete physical reasons.} and it is just such an independence which, on the one hand, would hypothetically provide it with full genericity, permitting its coherent operation at the different scales, and, on the other hand, would make it compatible with the reductionist view of causal influences from bottom features exclusively. The most demanding issue in our framework is the prefiguration of paths for complexity building we associate with the mode mixing possibilities of the generalized Landau scenario, and which would arise from phase space topological constraints that seem completely independent of any physical interaction. Less stringent, but also clearly unrelated to the physical interactions among matter constituents, are the alleged reasons for the asymmetry of likelihood between building and deconstruction of mode mixing, through which the evolutionary arrow would work. These two ingredients arise in the framework through the ontological view of the world as a deterministic dynamical system and after consideration of what things the dynamical systems can do. In this view, the physical world should be a particular dynamical system just describing a particular trajectory among its phase space possibilities, and both of such particularities should be determined by how the existing things and their causally interrelated properties have been at a certain moment, are just now and subsequently will be. Nevertheless, independently of such an existing reality, we expect that the given trajectory would exhibit generic features arising from the intrinsic constraints of what the dynamical systems can do, and our evolutionary framework rises like a tentative elucidation of such features under the guide of the innerly observable behavior of the world. \footnote{The evolutionary framework opens a way to tackle epistemological preventions against a scientific explanation of the whole world functioning, or against the thinker possibilities to explain thinking, by attributing common developing scenarios to such different things of the existing reality.} Other features we could associate with the intrinsic constraints are the reasons sustaining the robustness and persistence capabilities inherently and generically manifested by the emergent dynamical structures, possibly including the positive feedback we have supposed among the appropriate circumstances for structural evolution, or the suspected ways through which the developing complexity could influence on the evolutionary efficiency itself. Thus, our tentative description of the appropriate circumstances for oscillatory building contains elements that could be associated with intrinsic reasons, but it is clear that, although generically posed within the evolutionary mechanism, the presence and proper working of what we have called plasticity, as well as the actual occurrence of proper stimuli among interactive parts, imply conditions on the existing reality. That the existing reality effectively includes particularities fulfilling such conditions appertains to our fourth assumption for sustaining the evolutionary framework. Nevertheless, we peremptorily need to elucidate how this could happen in the supposed dynamical system of the world since, up to now, we have been arguing against the occurrence of structural evolution in mathematically defined systems, whether autonomous or non-autonomous.

Oddly enough, our reasoning requires to explicitly presuppose the entire universe just like a single and perfectly defined system, in which all its physical contents is involved in the common structure of dynamical interrelations, in which the causally related sequences of changes and transformations occur without room for arbitrariness, and in which the lack of environmental influences implies absolute autonomy. Thus, we are assuming strict determinism in the sense that everything would happen as strict result of previous things, without emergence (submergence) of anything from (into) nothing and, in addition, we are considering the world like a mathematically defined autonomous system in the sense that, at every moment, the consequent evolution would be strictly governed by the structure of dynamical interrelations. Under this abstract and holistic view, in which all the relevant details are taken into account, there is no true novelty but only strict consequence, and we don't need to solve the multifaceted mystery of self-organization as structural evolution by activity-induced plasticity effects with inherent persistence, but to explain features of the trajectory associated with the existing reality from the assumed nature of dynamical system for the whole world and from the assumed appropriateness of the actual system. Our evolutionary framework tentatively introduces the elements for one of such a kind of explanation, and its substantiation would primarily require better knowledge on the possible dynamical scenarios to deal with the world complexity and a (probably probabilistic) characterization of the occurrence of proper features in generic trajectories of such scenarios.

Nonetheless, the strict determinism of such a supposed world would look rather different when, according to our actual circumstance of observers, we consider either the whole or, more usually, parcels of the whole and try concrete descriptions by defining the relevant properties to characterize the effective interrelations underlying the dynamical systems behavior of a given subsystem in its environment. We establish in this way concrete phase spaces, far away from the abstract one we have assumed for the real world, depending on both the chosen scale and description accuracy, and whose applicability would be surely temporary. The generic feasibility of such a descriptive approach corresponds to our assumption (3). In any descriptive process, the necessarily neglected details, easily attributable to microscopic scales but possibly to higher levels also, imply strict determinism breakage at the observer level of description through two opposite kinds of perceptible effects: the ubiquitous presence of random noise and the more peculiar occurrence of structural evolution. The latter can be convincingly explained within the deterministic world view by attributing it to omitted details that, if properly incorporated into the description, would convert the apparently structural nature of the observed phenomena into a concatenated part of the current dynamical activity. Such details should include the euphemistic plasticity and external stimuli we have introduced as appropriate circumstances for structural evolution in Subsection 4.2. \footnote{Let us insist that structural transformation deals with changes in the organized structure of dynamical interrelations appreciated by the observer, not with behavioral bifurcations induced by quantitatively varying environmental properties by now included into the description. Nevertheless, the occurrence of novel behaviors can activate underlying plasticity effects through which the interacting subsystems can structurally transform at the eyes of the observer.} Notice, however, that such an explanation will be always incomplete and will successively require more and more details up to reach the abstract level of the dynamical interrelations strictly governing the real world, where structural evolution should have vanished. Alternatively, or perhaps equivalently, one could say that structural evolution does not admit mathematical description or, in other words, that a mathematical system cannot contain the reasons of its structural transformation or, perhaps better, that a (physical) system exhibiting structural transformation should be considered mathematically non-well defined at the level of the observing describer.

Within the deterministic view, the ubiquitous noise could be interpreted as arising from the unavoidable participation of huge numbers of omitted degrees of freedom sustaining relatively small but differentiated effects upon each one of the actually considered subsystem properties, provided that a justification for its apparent randomness could be generically argued. We find this explanatory way feasible and, since the same omitted details can be involved also in the apparent genesis of structural evolution, it is then relevant to ask to what extent the apparent randomness of noise could enhance the efficacy of the evolutionary processes in the sense of giving more chance to the occurrence of a given set of appropriate circumstances at the proper moment. On the other hand, at this point, as an alternative view, we could tentatively consider the possibility of an intrinsic source of randomization by supposing some extent of arbitrariness within the causal workings of the world. Before asking about the origin of that source we must realize its obligated physical nature and the consequent non-genericity of its participation. Curiously, arbitrariness cannot arbitrarily arise but needs a physical reason related to certain kinds of processes obeying definite rules, perhaps of quantum nature. Then, by taking into account at what scale, how and in what circumstances such processes introduce arbitrariness into the otherwise deterministic workings of the world, one should be able to justify three things: the broad background of random noise, the apparent preservation of a coarse deterministic evolution, to which the ordered working of the existing structural organization should presumably be attributed in some way, and the generation of new dynamical organization with cumulative and enduring capabilities in front of the randomization tendency, and all this for the different scales of observation. In this context, the genesis of structural organization would require some kind of causal ignition, i.e., sources of true novelty in the form of particularly fortunate physical events randomly arising from arbitrariness but properly altering the causal sequences within the otherwise preserved coarse deterministic evolution, and here is from where the proposals of order from disorder can emerge. At the best of our knowledge, however, a rationale covering such a conjunction has not appeared in the literature. \footnote{It is worth mentioning the stochastic resonance, which manifests how a non-oscillatory but biestable system embedded in a noisy environment is able to resonantly enhance its oscillating response to an external modulatory driving within a certain range of noise level because the mean time between noise-induced switching transitions depends on such a level In our view, this phenomenon, as well as other fluctuations-induced effects, does not represent emergence of order from disorder but the ocurrence of a kind of behavior in a system subjected to the pooled action of environment and driving. Simple variations in the driving parameters or in the noise level would imply the behavior vanishing, like happens in any externally-driven bifurcation of a non-autonomous system.}

Finally, a comment concerning quantum mechanics as a source of indeterminacy is worthwhile. The ontological reasons against determinism, arisen in philosophy of science mainly from its presumed inability to account for the appearance of novelty, seemed to be confirmed when quantum mechanics was widely accepted and the classical determinism constraint broken down, and, in this way, the position of other sciences became apparently more comfortable with physics. Nonetheless, it is worth stressing that the emergence of novelty persisted unexplained, without attempts relating it to any quantum indeterminacy, probably because the nature and consequences of such indeterminacy remain unclear within the unsolved interpretational problem of quantum mechanics \cite{87}. In fact, it is worth remarking that, even assuming the microscopic activity underlying everything as obeying quantum mechanics, it is today unclear whether physics is imposing or not any kind of indeterminacy in the functioning of the world. There are interpretations of the quantum theory maintaining strict determinism, while the dominant orthodox interpretation introduces causal indeterminacy exclusively through the wavefunction collapse it supposes to occur when properties of the corresponding quantum object are experimentally measured (and intellectually realized), but does not say anything at all about what happens within the parcels of reality free of scientific scrutiny. Thus, it seems clear to us that the most appropriate view for tentatively analyzing the running of the world is to assume its main streams as causally and deterministically sustained, as a first approximation at least and without excluding additional qualitative roles for undetermined events with probabilistic causation. At this respect, however, we don't find any reason to consider indeterminacy more able than determinacy in yielding the dynamical emergence of novelty required for explaining, for instance, the origin of life, the evolution of biological species, or the introspective sensation of creative and decisory thinking each one of us is continuously experiencing.

\section*{Acknowledgments}

The content of this paper is fruit of work devoted to the development and understanding of a family of physical devices, the BOITAL family, to which J. I. Rosell, F. Boixader, J. Farjas, R. Pons, and M. Figueras have also significantly contributed. The paper is dedicated to \textit{Design for a Brain}, a book by W. R. Ashby, which we have encountered just when ending the work and whose reading has convinced us that if the author had met the BOITAL devices, in addition to his Homeostat apparatus, we would have found the content of our paper in it. We are deeply grateful to one of the referees for useful and stimulating comments. Financial support from the Spanish and Catalan Governments, under Grants BFM2002-2366, FIS2007-66944-C02-01, FIS2008-06024-C03-02, SGR2001-187, and SGR2005-358, is acknowledged.

\appendix
\section{OPTIMUM SCENARIO FOR THE GENERATION OF COMPLEX OSCILLATORY BEHAVIOR}

We consider $(N,m)$ systems, as defined in Eq. (1), properly designed to posses full $m$-dimensional arrays of fixed points and to sustain Hopf bifurcations on these points, and we want to elucidate the optimum circumstances for achieving complex oscillatory behavior by nonlinear mixing of the different oscillation modes. The analysis develops essentially from generic considerations about the invariant manifolds connecting the various limit sets and the influences on such connections of the Hopf bifurcations of the fixed points, always trying to consider circumstances of structurally stable systems only. In two points of the analysis, our limited knowledge on the invariant manifolds behavior is surpassed by means of conjectures. Bifurcations of the periodic orbits, which surely occur as chaos probably appears, are considered secondary for the mixing mechanism and not taken into account. The presentation of the appendix presupposes previous reading of Subsection 3.1.

\begin{figure}[t]
\centering
\includegraphics[width=0.74\linewidth]{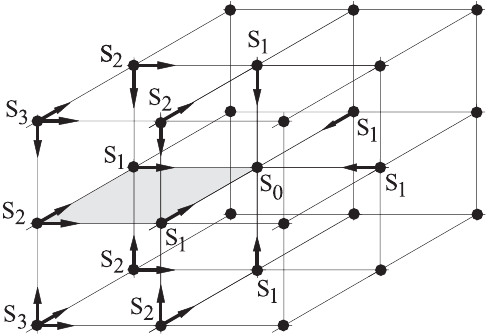}
\caption{Schematic network of saddle-node connections between the fixed points of the attraction basin of $S_{0}$ for an $N$-dimensional system with $m=3$. The shading denotes the two-dimensional unstable manifold of a $S_{2}$ point while arrows and thin lines mark the one-dimensional saddle-node connections between pairs of fixed points differing by one unstable dimension.}\label{FigA1}
\end{figure}

We first need to realize how the $m$-dimensional array of equilibria around a given point $S_{0}$ is structured, before the occurrence of any Hopf bifurcation, and how the unstable manifolds of the saddle points are organized like a sort of multidimensional Russian doll to built the basin of attraction of $S_{0}$. The array contains fixed points $S_{j}$ with the unstable manifold dimension $j$ varying from 0 to $m$, because the additional $N-m$ dimensions are assumed of common stability and attractive for all of the points, in order to acchieve structural stability and to guarantee the existence of an attractor, respectively. By extending to an arbitrary $m$ value what can be directly seen for $m$ up to 3 in Fig.~\ref{FigA1}, we deduce that:

\begin{list}{$\bullet$}{\setlength{\itemsep}{-4pt} \setlength{\topsep}{0pt}}
\item All the fixed points other than $S_{0}$ lie in the separatrix of the basin of attraction.
\item In the atraction basin, a given point $S_{j}$ is asymptotically connected to $\binom{j}{j'}$ points of type $S_{j'}$, with $j'$ varying from $(j-1)$ to 0, through the $(j-j')$-dimensional invariant submanifold describing the intersection of the unstable manifold of $S_{j}$ with the stable manifold of $S_{j'}$. The saddle-node connections are one-dimensional (1D) submanifolds connecting pairs with $j-j'=1$ and, in regular drawings like that of Fig.~\ref{FigA1}, the $(j-j')$D connection from $S_{j}$ to $S_{j'}$ should appear like a hypercube with the saddle-node connections as edges.
\item  On the other hand, the point $S_{j}$ receives connection from $2^{j'-j}\binom{m-j}{j'-j}$ points of type $S_{j'}$, with $j'$ varying from $(j+1)$ to $m$, through $(j'-j)$D submanifolds of its stable manifold.
\item  In particular, the number of points of type $S_{j}$ connected to $S_{0}$ is equal to $2^{j}\binom{m}{j}$ and, including $S_{0}$, this means a total number of $3^{m}$ fixed points in contact with the basin of attraction.
\end{list}

\begin{list}{$\bullet$}{\setlength{\itemsep}{-4pt} \setlength{\topsep}{0pt}}
\item Each $S_{j}$ is the saddle partner of $j$ saddle-node connections with different points $S_{j-1}$ and the node partner of $2(m-j)$ connections with different points $S_{j+1}$. Except for the $S_{1}-S_{{0}}$ connections, all the rest of $S_{j}-S_{{j-1}}$ connections are heteroclinic and lie within the separatrix, but all the saddle-node connections are structurally stable.
\item The network of saddle-node connections contains multiple pathways from a given $S_{j}$ to $S_{0}$ and such pathways superpose with those of other fixed points.
\item The jD unstable manifold of a given $S_{j}$ may be visualized by considering two steps: i) it emerges from the fixed point bounded by $j$ different $(j-1)$D submanifolds, one for each combination of $(j-1)$ of the $j$ saddle-node connections from $S_{j}$ to the neighboring points $S_{j-1}$, and ii) it is then bordered and collected toward $S_{0}$ by the unstable manifolds of the $j$ neighboring points $S_{j-1}$.
\item By applying the same reasoning to the unstable manifold of the $S_{j-1}$ and so on up to $S_{1}$, it is possible to imagine the hierarchical organization of unstable manifolds within the unstable manifold of a given $S_{\textit{m}}$.
\item The unstable manifolds of the various points $S_{m}$ form an ensemble of $2^{m}$ $m$D hypercubes around $S_{0}$, each one involving a total of $2^{m}$ fixed points.
\item The basin of attraction of $S_{0}$ is obtained by extending the structure of $m$D hypercubes with the additional $N-m$ attractive dimensions. The separatrix is defined by the $(N-1)$D stable manifolds of the $2m$ points of type $S_{1}$ and includes a number $j$ of $(j-1)$D unstable submanifolds of each saddle point $S_{j>1}$.   
\end{list}

\begin{figure}[t]
\centering
\includegraphics[width=0.45\linewidth]{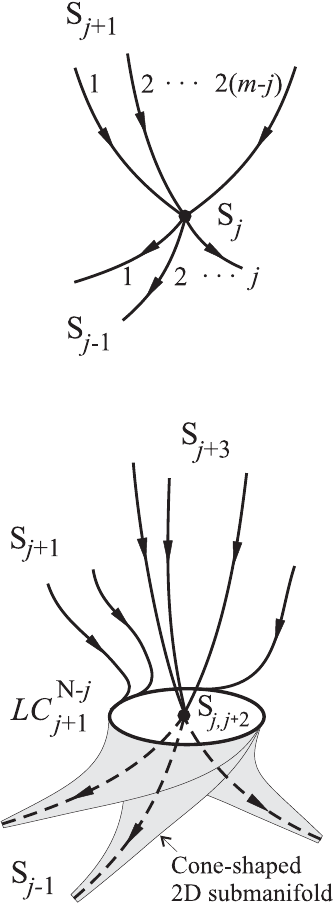}
\caption{Before and after the first Hopf bifurcation within the stable manifold of a fixed point $S_{j}$. Without depicting fully the invariant manifolds, the scheme illustrates: 1) How the stable manifold of the limit cycle LC collects the 1D saddle-node connections previously arriving to the point from neighboring $S_{j+1}$ points and how its unstable manifold incorporates 2D submanifolds connecting to the $S_{j-1}$ points of the next lower level. These are the basic ways of mode mixing through which the $j+1$ level will influence the $j$ level and the $j$ level will influence the $j-1$ level, respectively. 2) How the bifurcation can create a new set of 1D saddle-node connections arriving to the bifurcating point from neighboring $S_{j+3}$ points. These 1D submanifolds will be collected by the stable manifold of the second limit cycle emerged from $S_{j}$ and will provide for a direct way of influence from the $j+3$ level to the $j$ level. The depicted 2D submanifolds of LC do not cross one another because more than one of them have sense only if $j\geq2$ and $N\geq{j+2}$.}\label{FigA2}
\end{figure}

Secondly, we consider the occurrence of Hopf bifurcations and try to analyze what happens to the invariant manifolds connecting the various limit sets. We assume appropriate nonlinearities to sustain the emergence of a limit cycle from each one of the bifurcations. For simplicity, all the bifurcations are supposed to be supercritical and, for concreteness, the label $S_{q}$ for the fixed points is maintained unchanged after their bifurcation but the notation $S_{q,q'}$ is occasionally used to indicate both the original and actual unstable dimensions. To achieve optimum oscillatory behaviors we consider bifurcations occurring within the initially stable manifolds of the fixed points only:  

\begin{list}{$\bullet$}{\setlength{\itemsep}{-4pt} \setlength{\topsep}{0pt}}
\item Each fixed point $S_{j}$ experiences successive 2D oscillatory instabilities over pairs of stable dimensions up to exhaust its $(N-\textit{j})$D stable manifold, while the original unstable manifold does not participate in order to preserve the way of influence toward the attractor and neighboring saddle points of lower $j$ value. On the other hand, the unstable 2D submanifold of $S_{j}$ additionally created at a given Hopf bifurcation will be the source, at the next bifurcation, of a 3D submanifold connecting the new limit cycle to the previous one. Is through this submanifold that the two oscillation modes of the same fixed point mix one another by means of either a torus bifurcation on one of the limit cycles at the frequency of the other cycle or the emergence of localized bursts at one of the frequencies on the limit cycle of the other frequency. 
\footnote{We presume that both kinds of mixing can occur at the same time but alternatively in one or another of the two limit cycles. However, our reasoning is based on a few numerical results and we cannot be ascertain about how generic such a behavior may be and in what conditions the torus may appear on the first or second cycle (see Fig.~\ref{Fig1bis} and $^7$).} 
The mode mixing at the level of a fixed point may be rather general because the point will maintain a 2D submanifold connection to each one of the limit cycles emerged from it and then any new cycle will emerge with a set of 3D unstable submanifolds connecting it to each one of the previous cycles.
\item The first Hopf bifurcation of $S_{j}$ produces a limit cycle $LC^{N-j}_{j+1}$, if $j\neq0$, or $LC^{N}_{0}$, if $j=0$, where the superscript and subscript indicate the dimensions of the stable and unstable manifolds, respectively.
\footnote{If subcritical, the Hopf bifurcation creates a cycle $LC^{N-j-1}_{j+2}$, if $j<N-2$, or $LC^{0}_{N}$, if $j=N-2$, that will exist between the oscillatory instability and a previous cyclic saddle-node bifurcation, the node partner of which would be a limit cycle of the same type as that directly created in a supercritical bifurcation.}
The stable manifold of the emergent cycle collects, at least partly, the $(j'-j)$D submanifolds arriving to $S_{j}$ from $S_{j'>j}$ points
\footnote{This is because, after the bifurcation, $S_{j}$ will have a stable manifold with two dimensions less than the limit cycle and because the 2D instability may be generically expected to occur not orthogonal to any of the $(j'-j)$D submanifolds. For $(j'-j)>2$, the submanifold can split in two by maintaining a structurally stable connection of dimension $(j'-j-2)$ with the bifurcating point, as determined by the stable eigenvectors of this point.} and the various submanifolds approach the cycle by tangency under a well-defined organization. In this way the cycle is receptive to influence of the oscillation modes of higher levels of fixed points. On the other hand, for $j>0$, the unstable manifold of the emergent cycle appears as the $(j+1)$D border of the 2D-expanded unstable manifold of $S_{j}$, actually $S_{j,j+2}$. It describes a sort of multidirectional helical motion arising from the combination of the emergent oscillation, working in parallel to the instability plane, and the motion away from $S_{j}$ governed by its previous $j$D unstable manifold. In this way, the oscillation mode extends its influence for the phase space region around the unstable manifold of the limit cycle and, in particular, can approach the various fixed points $S_{j'<j}$ through the corresponding $(j-j'+1)$D submanifolds. We assume optimum mode mixing transport along the cone-shaped 2D submanifolds connecting the limit cycle to the next $j-1$ level (see Fig.~\ref{FigA2}), and this will be justified below as due to the possibility of homoclinic connections of the cycle based on such unstable submanifolds. Thus, in its turn, this means that the higher levels influence on the first cycle emerged from $S_{j}$ should arrive primarily from the $j+1$ level.
\end{list}

\begin{list}{$\bullet$}{\setlength{\itemsep}{-4pt} \setlength{\topsep}{0pt}}
\item Concerning the bifurcating fixed point, the 2D reduction of its stable manifold will affect the connections arriving to it from higher levels, as noted in $^{37}$. In particular, new 1D saddle-node connections can be formed between this point, actually $S_{j,j+2}$, and fixed points with a $(j+3)$D unstable manifold like those of the $j+3$ level (see Fig.~\ref{FigA2}). These connections are not generic but represent a possibility only and their number is limited up to $2(m-j-2)$. Similarly, after the second bifurcation, $S_{j}$, actually $S_{j,j+4}$, can receive 1D saddle-node connections from fixed points of the $j+5$ level, and so on for the successive bifurcations. In its turn, before doing any bifurcation and through its unstable manifold, $S_{j}$ can contingently establish additional saddle-node connections to fixed points of lower levels than the $j-1$ one, like a $S_{j-3}$ point after its first bifurcation, actually $S_{j-3,j-1}$ or a $S_{j-5}$ point after its second bifurcation, actually $S_{j-5,j-1}$. 
\item The second (third) Hopf bifurcation of $S_{j}$ produces a limit cycle $LC^{N-j-2}_{j+3}$ ($LC^{N-j-4}_{j+5}$) and, similarly to the previous bifurcation, the stable manifold of the cycle collects, at least partially, the unstable submanifolds arriving from upper levels to the bifurcating point. In particular, the new cycle collects the 1D submanifolds contingently arriving from $S_{j+3}$ ($S_{j+5}$) points and through these connections it will be particularly sensitive to mode mixing influence from that higher level, as a consequence again of the possibility of homoclinic processes. The new cycle may receive influence also from the second (third) cycle emerged from any of the $S_{j+1}$ points previously connected to $S_{j}$, as will be argued now while considering the influence toward lower levels. For concreteness, we consider the case of the second limit cycle: the unstable manifold of the second cycle appears as the border of the $2D$-expanded unstable manifold of $S_{j,j+2}$, actually $S_{j,j+4}$, and to elucidate toward where the new oscillation mode can effectively propagate we need details on how such an expansion works. Firstly, recalling the mode mixing at the $S_{j}$ level, we find reason to expect that the second mode oscillations incorporated to the first cycle can be transmitted along the $2$D cone-shaped submanifolds of this cycle toward the first cycles of the $j-1$ level and that this will happen through appropriate folding deformation of such conic submanifolds. Secondly, we need to realize how would be the $4D$ unstable submanifold of the second cycle arising from the $2D$-expansion of each cone-shaped structure and, in particular, if either it would remain integrally connected to the first cycle of the corresponding $S_{j-1}$ point or it would spread toward the fixed point or other limit cycles emerged from this point. In light of the numerical results for $m=1$ systems involving a $S_{1}-S_{0}$ saddle-node pair of fixed points, we conjecture that the unstable manifold of the second cycle of $S_{j}$ is generically able to connect to the stable manifold of the second cycle of $S_{j-1}$, and the same between the pairs of cycles of equal order. As will be shown below, a given cycle of the node partner of a saddle-node connection is compatible to become homoclinic orbit to the cycle of same order emerged from the saddle partner.
\end{list}

\begin{figure}[tp]
\centering
\includegraphics[width=0.55\linewidth]{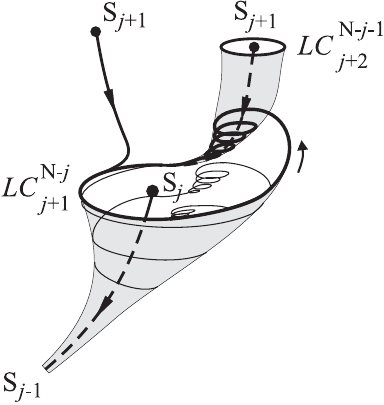}
\caption{Scheme illustrating two different things: 1) How the growing of the first limit cycle born from $S_{j}$ is affected by an ascending corkscrew around the unstable submanifold arriving to it from the first cycle of a point $S_{j+1}$, how in this way nonlinear mixing happens in the growing cycle and how the process is associated with the approach to a homoclinic connection of the higher cycle (stable manifold not drawn), and 2) How the influence of level $j+1$ on level $j$ can be transferred to level $j-1$ if the corkscrew upon the cycle is assumed to propagate along its unstable submanifold connecting to the lower level, so that each trajectory turn around that submanifold reproduces the shape and time features of the cycle. The depicted situation requires $j\geq{1}$ and $N\geq{j+3}$ and this allows for the corkscrew bending of the 2D cone-shaped submanifold.}\label{FigA3}
\end{figure}

Thirdly, we consider the parametric growing of the limit cycles and try to illustrate how the possibility of homoclinic connections within the structure of invariant manifolds determines efficient ways of oscillation mode mixing from higher to lower $j$ levels:

\begin{list}{$\bullet$}{\setlength{\itemsep}{-4pt} \setlength{\topsep}{0pt}}
\item The first limit cycle of a given $S_{j}$ begins by growing within the corresponding instability plane but it can subsequently deform along the unstable submanifolds arriving to it from $j'>j$ levels (see Fig.~\ref{FigA3}). The efficiency of such a deformation is associated with a process in which the growing cycle is approaching to become homoclinic orbit to a given saddle limit set of a $j'$ level that, in its turn, is going to form a homoclinic loop involving the unstable submanifold along which the growing cycle is deforming. This implies the condition that the $j'$ saddle set must have unstable and stable manifold dimensions compatible with belonging to the growing cycle, and this is only possible if $j' = j+1$, i.e., if there has been a saddle-node connection between the fixed points.
\footnote{It may be shown that, in the context of supercritical Hopf bifurcations, the first cycle emerged from the node partner of a saddle-node pair of fixed points can become homoclinic orbit to the saddle partner (before doing any bifurcation) or to the first cycle emerged from it, i.e., the cycle $LC^{N-j}_{j+1} (LC^{N}_{0}$ if $j=0)$ born from $S_{j}$ can become homoclinic orbit to any of the $S_{j+1}$ points saddle-node connected to $S_{j}$ or to the cycle $LC^{N-j-1}_{j+2}$ born from any of these $S_{j+1}$ points. The saddle-node connection between fixed points guarantees the manifold dimension compatibility and the existence of a submanifold connection driving the growing limit cycle toward the homoclinic process. The manifold dimension compatibility is also fulfilled between the cycles of same order emerged from the two fixed points, i.e., the second cycle of $S_{j}$, $LC^{N-j-2}_{j+3}$, can become homoclinic orbit to the second cycle of a $S_{j+1}$, $LC^{N-j-3}_{j+4}$, and so on. In this case the submanifold connection between cycles is not so obvious but, as discussed in the second part of the appendix, we expect it will happen naturally through the Hopf bifurcation generating the cycle of the saddle partner.}
In conclusion, we find reason to assume as generic, within the optimum scenario, that the parametric growing of the first limit cycle of $S_{j}$ is mainly occurring toward one (or several) of the saddle-node connected $S_{j+1}$ points and it happens in the form of a corkscrew effect sustained by the unstable manifold of the first limit cycle emerged from the $S_{j+1}$ point. Like shown in Fig.~\ref{FigA3} for the generic case and in Fig.~\ref{Fig1} for a $S_{0}$ point, the corkscrew works around the cone-shaped 2D submanifold ending by tangency on the growing cycle and, although its enhancement is associated with the approach to homoclinicity of the first cycle of $S_{j+1}$, the mixing at the level of this fixed point implies the transfer of all the oscillation modes of $S_{j+1}$ to the growing cycle.
\item The simultaneous growing of the limit cycle of $S_{j}$ toward two or more of the $S_{j+1}$ points (or their first limit cycles) implies that it is going to become a heteroclinic cycle connecting such saddle limit sets, provided that simultaneous tangency to the saddles would happen. Such a simultaneity is far from generic and its achievement usually denotes a proper symmetry into the system.
\item In generic circumstances, the growing cycle would be destroyed after tangency to one (or more) of the saddles and, although oscillation mode influences could remain in phase space, it does not seem convenient for optimum mode mixing.  
\item The second and successive limit cycles emerged from $S_{j}$ initiates their growing within the corresponding instability planes and they may be also affected by ascending corkscrews toward saddle sets of higher $j$ level to which can become homoclinic. Here again the 1D saddle-node connections between fixed points before the corresponding bifurcation indicate possible homoclinic connections and effective ways of mode mixing. Thus, the second cycle of $S_{j}$ can receive direct mode mixing influence from some of the $S_{j+3}$ points initially connected to $S_{j}$, the third cycle from some of the $S_{j+5}$ points, and so on. Nevertheless, such a kind of connections occur contingently only and more usual will be the mode influence from cycles of equal order of the $j+1$ level. As previously described, we expect that the cycle emerged from the second (or successive) Hopf bifurcation of a $S_{j+1}$ point will present a unstable submanifold connection to the second (or successive) cycle emerged from $S_{j}$ and, as noted in $^{38}$, the latter is compatible to become homoclinic orbit to the former so that an ascending corkscrew can affect the latter toward the former. We are unable to characterize how many of such connections can a given limit cycle receive but it is clear that some limit must exist.
\item It is worth noting that mode mixing at the level of a given fixed point is not directly associated with any homoclinic process but the invariant manifold interconnection between limit cycles plays a clear role also in this case. 
\end{list}

Finally, in view of the observed mixing efficiency in the $S_{1}-S_{0}$ connections of $m=1$ systems, we consider feasible the possibility of a chain of influences within the cluster of fixed points and limit cycles, and then we conjecture that:

\begin{list}{$\bullet$}{\setlength{\itemsep}{-4pt} \setlength{\topsep}{-2pt}}
\item Strong enough $(j+1)$ influences on a limit cycle of level $j$ can transmit along the unstable manifold of this cycle toward the next lower level, and so on. This means that the corkscrew bending on the 1D orbit of the limit cycle is able to propagate along its cone-shaped 2D submanifold and then the successive rounds of the trajectories around such submanifold reproduce the intermittent mode mixing of the cycle (see Fig.~\ref{FigA3}). In fact, we have already assumed something similar for the mode mixing at the level of the given fixed point.
\end{list}

In summary, the described scenario suggests a robust mechanism through which $N$ degrees of freedom could sustain a very complex but strictly well organized oscillatory activity based on the intermittent combination of a large number of characteristic modes of the system. From top to bottom in the $j$ scale \footnote{The mode mixing mechanism seems to be more general. For instance, we have detected traces of the oscillation mode of the second cycle of $S_{0}$ on the first cycle of $S_{1}$ (see Fig. 8a of \cite{16}). Although weak these traces mean a way of mode mixing influence from bottom to top in the $j$ scale. We find significant that the influencing and influenced cycles are of types $LC^{N-2}_{3}$ and $LC^{N-1}_{2}$, respectively, so that the latter is compatible in becoming homoclinic to the former, but in their origin they lack of invariant manifold connection, so that our kind of explanation for mode mixing would require the ulterior formation of such a connection.}, the oscillation modes emerged through Hopf instabilities within the stable manifolds of the different fixed points can be transported (mainly along the saddle-node connections of the array) toward the attractor originated from $S_{0}$ and can manifest in the corresponding time evolution. The multiplicity of connection pathways from a given $S_{j}$ to $S_{0}$ means that a given oscillation mode could influence the attractor in different locations and within different mode combinations. Under optimum circumstances, the maximum number of modes that could appear intermittently together in the observable time evolution is the number of possible 2D instabilities within the stable manifolds of the fixed points in contact with the basin of attraction. 
\footnote{In fact, additional modes might be even incorporated by means of gluing bifurcations creating bigger attractors over initially separated phase space regions. One of these global bifurcations would involve successive homoclinic connections of one of the $S_{1}$ saddles (more properly, the first limit cycle emerged from $S_{1}$) at the two sides of the separatrix and the resulting attractor can be seen as the gluing together of previously existing attractors \cite{95}.}
In addition to the asymptotic state of the attractor, the scenario contains a wealth of transient trajectories describing complex oscillatory evolutions according to their specific courses toward the attractor and it is easy to imagine an extraordinary variety of expectable possibilities among them.    

This scenario is what we have called \textit{generalized Landau scenario} for nonlinear oscillatory mixing. It is obvious that the higher the number of intermittently mixed modes the higher the number of required conditions on the system properties, and one could then become skeptical about the opportunity to find proper systems achieving the coexistence of fixed points, the occurrence of Hopf bifurcations in the various points, and the oncoming of homoclinic loops and/or heteroclinic cycles to enhance the corkscrew mixing between neighboring $j$ levels. One might suspect that the development of systems exploiting such possibilities at a large extent would have sense only by entailing some evolutionary procedure favoring it in the natural world, as that tentatively considered in Sect. 4. At this respect, the expansive growing capability of dynamical possibilities and the topological constraints on the invariant manifold development would provide the scenario with two of the necessary ingredients for any hypothetical management substrate of complex behaviors, i.e., the richness of possibilities and a basis for ordered functioning.

Notice, on the other hand, that the scenario does not require the system be on any homoclinic or heteroclinic attractor but it simply concerns the variety of possible global connections as an optimization mechanism of mode mixing. Previous works have shown how heteroclinic cycles can sustain intermittent bursting behaviors in models related to turbulence problems \cite{96,8}, neural networks \cite{98} and others \cite{99}. These models have proper symmetries allowing for structurally stable heteroclinic attractors \cite{101} and the intermittent bursts correspond to the alternate sequence of the direct trajectories connecting a succession of saddles and the oscillatory activities emerged around such saddles. Heteroclinic cycles occur also in the models for macromolecular evolution yielding the so-called hypercycles \cite{20} and may be present in models exhibiting chaotic itinerancy \cite{CIKiT}. Such heteroclinic connections would be regarded as highly degenerate in the absence of symmetries
\footnote{The sets of fixed points of these systems have structures of invariant manifolds different from the generic one we have described, and the disappearance of equilibria by deforming the system necessarily requires bifurcations of codimension higher than one.}
and we consider their occurrence as a particular situation of the described scenario.

\section{CONJECTURES ON THE ONSET OF TURBULENCE}

The wide agreement among physicists in assuming the Navier-Stokes equation as including the essentials for describing the motion of fluids is in contrast with the lack of understanding of the emergence of turbulent behavior in both fluids and equation \cite{103}. Within the nonlinear dynamics approach, having excluded the quasiperiodic route and low-dimensional chaos, the efforts are now focused on the high-dimensional spatio-temporal chaos, which might presumably be nothing but another name for some kind of turbulence in mathematical systems simpler than the Navier-Stokes equation and whose onset mechanism remains unclear also \cite{5,8}. An alternative explanation has been tentatively raised by showing that low-dimensional models describing boundary layers \cite{96,8} or plane Couette flows \cite{49,105,106} may evolve on heteroclinic attractors intermittently visiting the neighborhood of a sequence of saddles, around each one of which some oscillatory behavior has emerged. Nevertheless, the symmetries these models need in order to posses heteroclinic cycles arise mainly from those of the flow configuration and a more general mechanism would be required for explaining the ubiquity of turbulence in arbitrary configurations.

From our point of view, by trying to imagine how the oscillatory behavior of the generalized Landau scenario may be in the case of a spatially distributed system, we arrive to something having common qualitative features with the turbulent states of moving fluids \cite{107}. In our picture, there is a high-dimensional but single nonlinear oscillator exhibiting a composition of spatio-temporal structures of different characteristic scales, combined ones within the others with sudden intermittencies, discrete levels of similarity and generic recurrences at the various scales, all these both in space and time. Of course, how such a hypothetical behavior could fulfill the characteristic statistical properties of turbulent flows would be an important detail, but we expect that it need not be a problem in light of previous works showing how the Kolmogorov five-thirds spectrum can emerge in the time averaged behavior of models \cite{108} and experimental systems \cite{109} based on coherent structures. 

To substantiate our suspicion we need to verify the occurrence of two things in the (phase space of the) moving fluid: first, that there is a profusion of fixed points organized in a multidimensional array of saddle-node connections, and, second, that these points have sustained or are near to sustain a number of Hopf bifurcations. There are in the literature some pieces of evidence for the coexistence of fixed points in problems of fluid mechanics. From the experimental side, the Taylor-Couette flow has been particularly investigated and a large number of different stable steady states have been observed in the same fluid subject to the same geometrical and boundary conditions \cite{110,111,112}. The multiplicity of observed states has been explained by inferring the existence of intermediate saddle steady states \cite{111} but the experimental limitations in observing saddles imposes the analysis of mathematical models. A few numerical studies have been done with the Navier-Stokes equation for the circular \cite{113} and plane Couette flow \cite{114,115,116,49,Gib,Hal}. In the plane flow case, three-dimensional time-independent solutions were found \cite{114} in the form of a saddle-node pair of solution branches appearing independently of the initial laminar solution and it was then shown that the node appear stable up to perform a Hopf bifurcation \cite{117}. Particularly relevant for our analysis is the extensive numerical search done by Schmiegel \cite{49} in which, although restricted to two symmetry groups, more than forty stationary solutions for the velocity field were found. These fixed points appear with increasing the Reynolds number through a succession of saddle-node and pitchfork bifurcations and among them only one appears stable while the rest are saddles. The bifurcation diagrams of \cite{49} (Figs. 5.1 and 5.2) suggest that the ensemble of fixed points must be in a multi-dimensional array like those described in Appendix A. The first saddle-node bifurcation creates a $S_{0}-S_{1}$ pair accompanying the initial $S_{0}$ laminar state with the stable manifold of $S_{1}$ as separatrix. Successive pairs of fixed points must appear with at least one $1$D saddle-node connection to previously existing points, if structural stability should be maintained, and the lack of any other stable state suggests that they built a multi-dimensional array around the basin of attraction of the new $S_{0}$ point. The eigenvalue spectrum reported in \cite{49} (Fig. 6.2) shows that the new $S_{0}$ point experiences a succession of (up to seven) Hopf bifurcations and such an abundance suggests that the other fixed points should experience Hopf bifurcations also. Recently, another extensive numerical study done at really high spatial discretization resolution has confirmed the occurrence of a large number of fixed points and has detected heteroclinic connections between some pairs of such points \cite{HalPhD,Gib,Hal}, in generic agreement with the multi-dimensional array of the generalized Landau scenario. The eigenvalue spectra reported in Appendix A of \cite{HalPhD} show how, for a given Reynolds number, the majority of fixed points have several complex eigenvalue pairs with positive real part, suggesting the generic occurrence of Hopf bifurcations in the equilibria. In fact, although the origin through Hopf bifurcations has not been examined, a large number of unstable periodic orbits have been detected by the same authors and they propose a visualization of the turbulent flow as a sequence of close passages to such orbits \cite{Cvi}. Other examples of coexisting steady states have been also numerically found in a diverging channel \cite{118} and after a sudden expansion \cite{119}. It is also noteworthy that the finite-dimensional models \footnote{See \cite{8}, chapters 4 and 9.} derived by Galerkin projection of the Navier-Stokes equation usually have multi-directional nonlinear vector fields, like that of Eq. (1), with $m$ equal to $N$ and with each nonlinear component depending of almost all the variables, so that these systems may potentially possess extended arrays of saddle-node connected fixed points. For instance, bifurcation diagrams with hundreds of equilibria, mainly saddles, have been found in a reduced $(N=19)$ model for the plane shear flow \cite{48}. A Galerkin spectral method has been also used for extensive numerical analysis of Rayleigh-B\'{e}nard convection in a cubical cavity and several equilibria and Hopf bifurcation points identified \cite{Puig}.

	 The involvement of Hopf bifurcations in the first observable steps of the turbulence transition is clear for a variety of flows, like the bluff body wakes and Taylor-Couette systems, although it is not so obvious for other flow configurations \cite{107}. In any case, the turbulence appears always abruptly, either from an oscillatory or from a static state, and without allowing clear distinction of more discrete events in the transition. This fact and the apparently contradictory one of the variety of known transition scenarios are two qualitative issues that any tentative explanation of the turbulent transition needs to deal with.

	In the Eulerian specification, the fluid in continuous flow for a given spatial region may be considered like a dynamical system with the phase space associated with the velocity vector field $u(r,t)$ and, if necessary, with other fields describing variable properties, like the pressure scalar field $p(r,t)$ (see $^{17}$). The dynamical mechanisms of mechanical nature are contained in the Navier-Stokes equation:
\begin{equation}
\frac{\partial{u}}{\partial{t}}+\left(u\cdot\nabla\right)u=-\frac{1}{\rho}\nabla{p}+\nu\Delta{u},
\end{equation}

\noindent
where the inertia field $\left(u\cdot\nabla\right)u$ expresses locally and instantaneously what force per unit of mass the moving fluid particles would need to adjust their velocity to that defined by the velocity field of the given moment. The excess or defect of actual force with respect to the required inertia determines the rate of local change in the velocity field or, in other words, such a difference characterizes the dynamical system. The three terms depend on the velocity and therefore sustain feedback. The quadratic dependence of the inertia term represents an obvious local source of nonlinearity. The viscous force, $\nu\Delta{u}$, is determined locally and linearly by the velocity field; it depends on relative velocities of neighboring positions and, unlike standard friction forces, it may have an arbitrary direction with respect to the velocity vector. The pressure force works always against the local pressure gradient, $\nabla{p}$, but the implicit relation between pressure and velocity fields hides dynamical interactions extending crosswise the fluid (i.e., Eq. B.1 must be properly supplemented to describe the dynamical system fully). In the ideal case of incompressible fluids $\left(\nabla\cdot{u}=0\right)$, the pressure and velocity fields instantaneously adjust one another, and their static relationship may be explicitly shown to be nonlocal and quadratically nonlinear. \footnote{The pressure at a given place is influenced by local properties of the velocity field from everywhere in the flow, such an influence decreases with the distance and is related to a quadratic combination of spatial derivatives of the velocity components. See \cite{121}, p. 30.} Complex feedback circuits arise from such velocity field interdependences and, in particular, from the mixing of velocity vector components by the two nonlinear terms. In addition to the constant fluid properties, $\rho$ and $\nu$, explicitly appearing in the Navier-Stokes equation, the parameters of the dynamical system lie in the boundary conditions and, in particular, they include the flow strength control usually characterized by the Reynolds number, $Re$. A practical view about the dynamic role of the boundary conditions is that they restrict and determine the phase space region where the system can move by describing trajectories according to the dynamical equation.
 
A fixed point of the flow is any velocity field that, being in accordance with the boundary conditions, maintains a local equilibrium between the actual total force and required inertia everywhere in the field, and this applies to both stable and saddle points. For small enough $Re$ numbers, the low velocities make the nonlinear terms irrelevant and there is the single fixed point derived from the rest state of the strictly linear system $Re=0$, which is surely stable. Increasing $Re$ means displacing the phase space region where the system may be towards higher velocities and, in this way, the nonlinearities become more and more relevant in the feedback circuits within the velocity field. We are not able to analyze such circuits in detail but find reason to conjecture that they intrinsically contain proper nonlinear dependences and competing effects for developing the ingredients of the generalized Landau scenario in flows of arbitrary geometries and boundary conditions. Thus, we presume that with increasing $Re$ what happens in general is that: 

\begin{itemize}\setlength{\itemsep}{-4pt} \setlength{\topsep}{0pt}
\item[i)] The fixed point of the initially at rest state moves in phase space along a non-straight line by reorganizing the velocity field and accordingly transforming the associated spatial pattern. \footnote{A peculiar exception is the initial fixed point of the plane Couette flow, which moves along a straight-line without spatial pattern changes because in it the three terms of the dynamic vector field remain equal to zero always and, in this way, the boundary-induced velocity growing can be adiabatically done without nonlinear influences. Similarly, the initial fixed point of the plane Poiseuille flow develops the parabolic velocity profile without nonlinear influences because in that velocity field the inertia term vanishes and the pressure force is just equal but opposed to the linear viscosity force.} This nonlinear transformation may occur without necessarily requiring any bifurcation of the equilibrium point, although it can of course become unstable and give up a new stable state, either stationary or oscillatory.
\item[ii)]	Additional fixed points appear through successive saddle-node or, more rarely, pitchfork bifurcations and their number indefinitely increases. The stationary velocity field of each fixed point corresponds to a peculiar static spatial pattern, non-necessarily regular but presumably containing steady structures of varied sizes and shapes in order to fulfill the boundary conditions. The majority of the new states are saddles, they appear in the phase space within a multi-dimensional array of saddle-node connections and their pattern characteristic lengths become presumably shorter with increasing the unstable manifold dimension. \footnote{But the pattern amplitudes not need to become smaller or, in other words, the successive equilibria do not correspond to a sort of perturbation series.}
\item[iii)] A succession of oscillatory instabilities takes place in the various fixed points (perhaps with exceptions due to particular symmetries) and we suppose that such bifurcations happen within the stable manifolds of the points. Each limit cycle describes a time-periodic velocity field with a spatio-temporal pattern derived from the static one of the corresponding fixed point. The cycle orientation in phase space (see $^{17}$) reflects the different amplitude of the velocity oscillations over the physical space. The oscillation frequency is not associated with any vortical or helical motion but with the velocity pattern changes propagating like a wave over the tortuous flowing medium, usually at a much smaller velocity, and its value does not arise from a typical boundary-value problem but is determined by the underlying competition loop in the dynamical interrelations. \footnote{One observable example is the stable limit cycle of the K\'{a}rm\'{a}n vortex street on the cylinder wake.}
\item[iv)] Effective nonlinear mixing between different limit cycles occurs to some extent and it sustains a chain of influences within the array of limit sets that can manifest upon the existing attractors, i.e., the system is developing a generalized Landau scenario. Let us remark our lack of evidence for this part of the conjecture in fluid flows, other than the turbulence itself and our belief that mode mixing might tentatively explain it. In the physical space, mode mixing means that the time evolving velocity field of the oscillation mode associated with the influenced cycle makes the feedback circuits intermittently effective in sustaining the oscillation mode of the influencing limit cycle, in different moments at different places. In this way, the velocity pattern wave of one cycle incorporates spatio-temporal features of the other.
\end{itemize}

While the variety of fixed points, limit cycles and transient trajectories of the nonlinear mixing scenario describe potential behaviors under the given boundary conditions and for different hypothetic initial values, the velocity vector field of the fluid flow is actually evolving asymptotically near one of the attractors, by exhibiting the corresponding spatio-temporal evolution to which we presumably attribute features of the turbulent behavior. \footnote{The fixed points and limit cycles, either stables or saddles, correspond to steady laminar flows of the whole fluid, and this applies also to periodic orbits affected by nonlinear mode mixing, in which, however, laminarity would apparently become blurred by the intermittent combination of a successively increasing number of different spatial and temporal scales. In addition to the high degree of mode mixing, the observable attractor would probably include chaotic features and noise effects.}
The observable transition to turbulence would be that associated with the transformation of the attractor where the system is actually evolving. The geometry and boundary conditions determine the details of the bifurcation diagram as a function of $Re$, mainly the relative order and accumulation extent of the two kinds of bifurcations, and, in this way, they define the peculiar transition features of each flow configuration. For instance, in the plane shear flow, the initial fixed point remains always stable but, while the other fixed points appear and presumably do Hopf bifurcations, it gradually becomes more sensitive to perturbations due to a reduction of its basin of attraction and the system easily switches to the other side of the separatrix, where a generalized Landau scenario has been presumably developed around the attractor emerged from the stable node of the first saddle-node bifurcation. More in general, the initial state, after significantly transforming its static pattern or after becoming a new stationary state through a pitchfork bifurcation \footnote{In the circular Couette flow, the first static state with Taylor vortices mathematically appears through a pitchfork bifurcation although, in experiments, the lack of perfect symmetry usually avoids the pitchfork and the state transformation happens without instability but with a saddle-node bifurcation nearby (imperfect pitchfork).}, makes also a Hopf bifurcation and the observable transition begins with a periodic state, perhaps followed by some quasiperiodic and low-dimensional chaotic states \footnote{As sometimes observed in the Taylor-Couette apparatus \cite{112}.}, perhaps followed by an intermittent two-frequency combination \footnote{As may tentatively be appreciated in the development of the K\'{a}rm\'{a}n vortex street in the cylinder wake.}, perhaps suddenly followed by the turbulent behavior. In any case, the accumulation of Hopf bifurcations and the consequent abrupt development of the oscillatory scenario would presumable explain the vigorous onset of turbulent states after a few distinguishable steps. There is also the possibility of attractor destruction through a homoclinic or heteroclinic bifurcation and then the oscillatory scenario would sustain a transient turbulent behavior while the system is evolving toward another attractor. In any case, it is worth stressing that the oscillation modes of the scenario would not correspond to the rotational motions of eddies and vortexs but to the variety of velocity pattern waves  appearing in a nested structure of spatio-temporal structures.

In conclusion, we are here conjecturing that the fluid flows and the Navier-Stokes equation (with boundary conditions) are formidable dynamical systems, possessing and being able to exploit large oscillatory capabilities through generalized Landau scenarios of high $N$ and $m$ values. We also conjecture that this is so because their structure of feedback interdependences contain proper nonlinearities and competitions, as well as because their usual control parameter activates the nonlinear mechanisms in a continuously increasing way that develops the Landau scenario properly. \footnote{Like $\mu_{C}$ in Eq. 3, but unlike what is usual in the investigation of dynamical systems, $Re$ acts as an effective scale factor on all the dynamical nonlinearities of any fluid flow.} 
We consider confirmed the scenario involvement in the plane Couette flow \cite{49,Gib,Cvi} but, of course, the conjecture validation needs to verify that there is no flow configuration yielding turbulence without having a profusion of fixed points doing Hopf bifurcations.

Finally, let us remark that the behavior of a real fluid surely involves more physical effects than those contained in the Navier-Stokes equation and, although in proper circumstances some of them may enrich the feedback structure of the system, such effects mostly act in a random way and introduce dynamically perturbing noise, surely larger than the numerical noise of the equation calculations. This suggests us that the conjectured Landau scenario would develop in the Navier-Stokes equation with richer features than in real fluids, primarily implying an extremely high number of fixed points and consequent effects upon the equation behavior.

\section{HYPOTHESIS FOR A DYNAMIC BRAIN}

The time evolutions of a variety of signals derived from living brains exhibit oscillatory patterns with differentiated characteristic features depending on the location and extension of the recording area, on the kind of recorded signal and the filtering procedure employed, on the environmental conditions stimulating sensory inputs, and on the particular moment at which the brain is found within its wake-sleep cycle \cite{40,131,132,133,134}. In mammalian brains, the oscillation frequencies cover from approximately 0.05 Hz to 500 Hz and a number of peculiar spectral bands are commonly distinguished. Certain oscillatory rhythms are observed at the cellular level, while others seem proper of the collective activity of neuronal ensembles. The recorded signals usually show different frequency rhythms, mostly combined in an intermittent way suggesting slow phenomena modulating faster ones. In fact, the wake-sleep daily cycle with its sleep hourly phases may be considered as a complex oscillation that modulates the activity of the whole brain.

The relevance of the brain oscillatory activity for its superior functions is not unanimously accepted among neurologists, but such a view is becoming dominant and some main ideas are now rising up \cite{131}. The first one is that the brain itself, without requiring sensory inputs, sustains an endogenous oscillatory activity \cite{136,132}. Secondly, the externally evoked neural activity represents the triggering and modulation of the endogenous dynamics by the sensory input, rather than directly reflecting the structure of the input signal itself \cite{137,138}. Thirdly, both the endogenous and the externally evoked activities occur with a certain degree of coordination across more or less extended spatial regions, as manifested by the electro- and magneto-encephalograms themselves, which express composite signals of undefined tissue volumes, and by the observation of phase synchrony between separated focal sites, of well defined spatio-temporal patterns, and of correlation between the activity of a single neuron and the spatial pattern where it is embedded \cite{139,140,141,142,143}.

The large-scale neuronal cooperation suggests a dynamical view of the brain that has been incorporated in a number of theoretical proposals relating consciousness to oscillatory dynamics \cite{144}, from which we selectively refer here some informative presentations \cite{145,146,147}. These models assume highly parallel and distributed information processing based on dynamical oscillations and associate the binding of percepts either with extended in-phase synchronization of neuronal oscillations at a particular frequency (40 Hz) \cite{131,140} or with coordination among the variety of oscillatory activities \cite{133,145,147}. Nevertheless, the models are unable to specify what kind of oscillatory behavior is underlying the brain operation and this must be related to the lack of knowledge in nonlinear dynamics about potential scenarios for such a complex oscillatory activity. The dynamical view has been also introduced in neural computation, as alternative to paradigms based on attractor neural networks, through approaches using the inherent transient dynamics of high dimensional systems under a variety of concepts: winnerless competition \cite{98}, chaotic itinerancy \cite{TSU} or liquid state machines \cite{151}.

The neuroscience oscillation-based proposals, together with the commonality between the oscillatory activity described in the electrophysiology literature and what we imagine for a generalized Landau scenario, induce us to consider the possibility of such a kind of scenario for sustaining a theoretical framework of the brain operation. Our analysis develops within the abstract context of the dynamical systems, with a sequence of  considerations about how the oscillatory activity could provide for some of the basic operative functions of a brain. The hypothetic dynamical system, that we call dynamic brain for concreteness, is firstly assumed with fixed structure and parametric properties, its oscillatory activity in the absence of any sensory input is related to the brain endogenous activity, and then, in the presence of sensory input, the externally-induced but intrinsically-governed activity is considered to provide a way of interactive processing between the information arriving from the environment and that stored into the system. As a second stage, learning and memory are introduced by supposing activity-driven plastic mechanisms upon certain parametric properties and such mechanisms are assumed properly regulated by a generic criterion directly connected to the oscillatory behavior. The main features of this hypothetical dynamic brain seem to accommodate well to the neuroscience views relating the brain operation to the coordinated activity of multiple brain oscillations \cite{133,145,147}.

\subsection{Tentative description of the dynamic brain}

We consider the dynamic brain as a high-dimensional dynamical system able to sustain states of a high degree of oscillatory instability, like those we associate with the generalized Landau scenario. Any operative function of this ideal brain must be exclusively based on the dynamically regulated activity of its variable properties. 

When trying a comparison with the living brain, we are unable to ascertain which and at what spatial scale the dynamically relevant variables would be. They might be either at the intracellular, cellular or cell assembly levels and, in fact, the dynamical mechanisms could involve the interrelation of properties associated with different scales. Consequently, we cannot identify the interdependences sustaining feedback loops and introducing competition and nonlinearity, as well as we cannot describe a correspondence for the fixed points. It is here worth evoking the fixed points of a fluid flow  (Appendix B) in order to avoid the association of a fixed point with just the rest state, and if, for concreteness, we imagine the dynamically-relevant brain variables as describing average activities of a number of properly chosen neuronal assemblies, and the relevant parameters not as synaptic properties of single cells but certain averaged properties characterizing the global influences among such assemblies, a fixed point would then be any dynamical equilibrium state in which all the relevant assemblies would maintain their mean activities constant in time. Presumed nonlinearities in the feedback circuits would provide for the multiplicity of equilibria, each one of them associated with a peculiar static pattern of mean activity levels for the relevant assemblies. By distinguishing assemblies of excitatory and inhibitory neurons one might introduce the required competition for oscillatory instabilities. \footnote{Different degrees of freedom not need to be based on spatially separated properties and effective competition between spatially coexisting properties can occur if they introduce opposite influences with different dynamic times in the feedback circuits. For a physical example, see \cite{65}.}
The system would not need to be or to have been on any fixed point, and it would be evolving in association with an attractor that extends toward phase space regions where oscillatory motions have appeared in relation to the Hopf bifurcations of the various equilibria. In other words, the generalized Landau scenario would provide for the coexistence of oscillatory possibilities associated with different configurations of neuronal assemblies and for a dynamically organized mixing of such oscillations.
\footnote{The number of oscillation modes could be really high even for relatively modest $N$ and $m$ values because of the exponential growing with $m$ of the number of possible fixed points. Recall also that each mode would be characterized by its oscillation frequency, which may be the same for different modes, and also by its peculiar influence on the different dynamically-relevant variables according to the limit cycle orientation in the phase space. Since such variables would correspond to differently localized properties, each oscillation mode should then exhibit peculiar spatio-temporal features. The nonlinear mixing would express how some modes combine with others along the different trajectories and, in particular, on the attractor.}

Finally, it is worthwhile remarking how varied (and difficult to interpret) may be the external observation of any spatially-distributed dynamical system having developed a generalized Landau scenario to some extent, like we assume for the dynamic brain, and for which the spatio-temporal features of turbulent fluids might provide, according to our presumptions, a tentatively useful insight. Recall that, in general, the number of variable properties of a dynamical system may be arbitrarily large with respect to the number of degrees of freedom, that each variable describes its peculiar time evolution, that any detection procedure defines in fact its own variable, that the detected signals may correspond to larger scales than those of the dynamically-relevant variable properties but also to shorter scales, that observing the relative behavior of two or more variables is just another way of introducing new variables, and that the spatio-temporal signal obtained from the simultaneous detection of a number of spatially-differentiated variable properties is peculiar of each spatial projection.

\subsection{Dynamic brain disconnected from sensory inputs}

In the absence of sensory input, the dynamic brain is evolving in the oscillatory state endogenously determined by its structure and parametric properties, which are assumed to remain fixed. Each one of its variable properties describes a peculiar sequence of oscillations with intermittent dominance of different characteristic frequencies and with the common feature for all the variables that the complex sequence repeats almost equal every longest characteristic period of the system. We envisage a generalized Landau scenario with the oscillatory activity essentially developed toward one (or a few) of the corners of the basin of attraction, so that the longest cycle would manifest with a part of strong oscillatory activity followed by another of relative calm, like in the low-dimensional example of Fig.~\ref{Fig8}. Such a kind of behavior suggests a sort of wake-sleep cycle in which calm would correspond to attentive waking and strong activity to sleep. In the wake stage, the major part of (dynamically-relevant) variable properties of the disconnected dynamic brain would remain almost fixed and only those related mostly to the slowest mechanisms would significantly vary. The lack of oscillatory activity implies the absence of operative functions and the waking system would remain strictly attentive to an apparently silent environment. The evolution of the slowest variables would drive the system to become asleep and then the rest of variables initiate their oscillatory activity according to the corresponding sequences imprinted in the attractor of the given Landau scenario. The slowest modulation of the sleeping activity would affect the whole set of variables and could then be associated with the succession of sleep stages within the wake-sleep cycle. In fact, we find feasible the design of scenarios with proper cyclic sequences of activity and calm to fit the different styles of wake-sleep cycles among animal species.

Of course, the interaction with the environment and, particularly, with other brains, is an intrinsic feature of any living brain and the disconnected brain must be considered as an element of prospective analysis for the dynamic brain. While the occurrence of sleeping activity may reasonably be expected independently of external influences, because the sleep features include a sort of disconnection, we cannot ascertain what cognitive actions a hypothetically disconnected brain could do when awake. For instance, according to our own introspection, we would be tempted to expect a continuous sequence of thoughts even in the absence of sensory input. In contrast to this, our framework indicates that the disconnected dynamic brain could not sustain cognitive experiences when awake and this fact induce us to interpret the line of thoughts of the waking brain as an externally-induced transient activity that would attenuate up to disappear if the lack of sensory input would continue for long enough time (perhaps several cycles). This is a consequence of the kind of phase space scenario we have assumed for the dynamic brain and, although scenarios sustaining continuous oscillatory activity are also possible, we exclude them because are not appropriate for the overall framework.

\subsection{Transient activity induced by sensory inputs}

In the phase space of the generalized Landau scenario, the flow of trajectories of the nonlinear mode mixing extend far away from the attractor and contain a wealth of different oscillatory patterns that, as pointed out in subsection 3.3, can be transiently induced by properly displacing the state of the system from the attractor. The flow richness around the multi-dimensional structure of invariant manifolds provides the system with a potential tool for identifying the external stimuli according to their correspondence with the induced characteristic transients and this is the way through which our framework introduces sensory input processing in the dynamic brain.

In the absence of sensory input, the dynamic brain is evolving on the attractor and we now consider the occurrence of a quick sensory input by simply supposing it like a sudden perturbation of a definite set of dynamically relevant variables according to the sensory input map. The state of the system has been suddenly displaced from the attractor by the perturbing vector, and then the intrinsic dynamics will determine the transient activity of asymptotic return toward the attractor. Such a transient will correspond to a single trajectory and will therefore describe a coordinated activity of the whole system, affecting certain variables more than others and with characteristic oscillatory patterns, all this according to the initial point of the trajectory defined by both the input map vector and the system state just before the perturbing moment. In the absence of plasticity effects, such a transient would be the exclusive response of the system to the sudden input and, by trying to associate it with basic brain functions like the experience of sensations and generation of motor answers, one might deduce requirements for the oscillatory properties of the assumed scenario. Without pretending to analyze the issue in detail, we find worth remarking some illustrative points. For instance, to achieve a defined enough relation between sensation and external stimulus, the induced oscillatory transient must contain a (probably fast) component fulfilling two conditions: i) to specifically characterize the given stimulus in relation to others, and ii) to be relatively independent of the system state at the perturbation moment. The example of Fig.~\ref{Fig8} suggests that the generalized Landau scenario can provide for such features when the system is evolving in the calm region of the attractor. This is in fact the main reason for choosing the kind of Landau scenario we have assumed for the disconnected dynamic brain. On the other hand, if the sudden input would correspond to a given scene momentarily projected on the retina, the induced transient trajectory would then be able to sustain the variety of sensations involved in the scene with enough extent of simultaneity. Furthermore, there is in fact a continuous stream of sensory input and the single trajectory describing the continuously reactivated transient should be able to provide for two complementary things: the individual identification of stimuli and the sequential combination of successive stimuli into the experienced sensation. This requires that the transient activity induced by a given stimulus must include oscillatory components independent of previous stimuli and oscillatory components dependent on such stimuli. The problem could be addressed by considering the different decay times of the various oscillation modes in the transient return to the asymptotic state and supposing that fast decaying components are induced with features independent of the slow decaying components remaining from previous stimuli, while slow decaying components are induced depending on the remnant activity at the excitation moment. In this way, one stimulus after another could sustain both the sequence of their individual sensation and a global sensation describing the sequence of a number of stimuli. The sequential excitation problem has been previously considered in proposals of frameworks for neural computation based on external perturbations of high-dimensional dynamical systems \cite{151,152}. Particularly related to ours is the framework based on the winnerless competition principle \cite{98}, in which the sequential memory is encoded to the heteroclinic connection of a network of saddle fixed points with one-dimensional unstable manifold \cite{152}.

We expect that the phase space flow of the generalized Landau scenario may be rich enough to provide specific transient trajectories for a representation of the variety of temporal sequences and spatial arrays describing the sensory world. The critical point would be the excitation ability to displace the state of the system to the proper place, in order to transiently induce oscillatory activity specifically correlated to the spatio-temporal pattern of a given external stimulus and with relative independence of previous stimuli and of the given moment within the waking stage. The analysis is complex because the excitation processes cannot be simply considered as a sequence of sudden vector displacements within an unperturbed phase space and their description must include details of sensory systems and their connectivity structure to dynamically relevant properties of the brain.

\subsection{Learning and memory}

In neurology, it is commonly assumed that learning and memory are achieved through activity-induced plastic changes and a wealth of plastic mechanisms have been found in biological brains, essentially at the level of synapses and in a variety of forms and time scales, but also in the excitability of neurons and the morphology of both cells and circuits \cite{153}. Plasticity may be regarded as an inherent feature of the nervous systems, potentially playing an essential role also for the brain development and repair and for its phylogenetic evolution. Nevertheless, there is a lack of detailed knowledge about how the plastic effects can properly participate in sustaining any of the brain functions, probably because it would imply the basis of how the brain works.

Learning through plasticity would require the occurrence inside the system of i) externally-induced activity specifically describing the information to be learnt and ii) activity-driven changes that, remaining after the stimulus disappearance, would provide the system with ability to identify next arrivals of the given stimulus by properly associating it with other memories. Thus, the processes of acquisition and storage must be able to introduce associative links with prior memories, but without excessively altering their content and retrieval. Furthermore, the achievement of memory stability during the life span in a so continuously stimulated plastic medium would seem unattainable without proper consolidation mechanisms. A theory of learning must be compatible with all these features and, therefore, it needs to imbricate the acquisition, storing, linking, consolidation and retrieval of memories to common functional mechanisms presumably including the essence of the brain operation. This is what we are trying to achieve in the dynamic brain framework and our aim is to discern the conditions required for it.

Tentative explanations of brain learning have been done by associating it with a given regulatory rule for the plastic mechanisms and supposing such a rule able to produce networks of interconnected cell assemblies specifically correlated to the different sensory inputs. Hebb \cite{156} predicted one of such rules by relating synapse strengthening to the simultaneous activity at both synaptic sides and, in fact, Hebbian and anti-Hebbian rules have been linked with long-term potentiation and long-term depression, two opposite kinds of lasting effects for which a large body of physiological and biochemical data exists \cite{157}. Nevertheless, learning under synapse specific rules implies stability problems and more complex rules including regulatory plastic mechanisms at the cellular or circuit levels have been proposed \cite{158,159}.

In the context of the dynamic brain based on states of high-degree of oscillatory instability, we find natural to tentatively associate the learning capability with an inherent tendency towards the effective incorporation of new oscillation modes into the system dynamics, i.e., novel information in sensory input excites oscillation modes previously absent in the intrinsic dynamics and the plastic mechanisms tend to incorporate them into the phase space by properly modifying a set of parameters. Such an inherent tendency would imply two different but related types of requirements for the dynamic brain, concerning the early appearance in the phase space flow of a previously absent oscillation mode and the learning ability of the plastic processes when the oscillations are there, respectively. The initial presence of a spiral oscillatory trace might be considered as a prerequisite for effective learning of the corresponding mode and it seems clearly related to the previous acquisition of other oscillation modes to which the new one could be linked. This suggests the building nature of learning and would associate it with the development of the generalized Landau scenario, concerning both the appearance of fixed points and the occurrence of Hopf bifurcations. By supposing small spiraling traces of a given mode already present in the system phase space, the consolidation of this mode into the nonlinear mixing would be enhanced by the first sensory input properly displacing the system state to excite such an activity if, for {\em appropriate circumstances}, positive feedback occurs between the mode activity and the plastic processes induced by it, i.e., if the plastic changes approach the system to the corresponding Hopf bifurcation or, more precisely, to the optimum presence of the oscillation mode in the phase space region accessible to the given sensory input. Thus, we might introduce learning capability in the dynamic brain by assuming plastic processes that, independently of the variety of their specific regulatory rules, work as a whole to consolidate the oscillation mode inducing them.

Let us illustrate how the inherent tendency to incorporate oscillation modes could provide our ideal dynamic brain with the features we are searching for. By assuming the generalized Landau scenario as potentially capable for representing the repertoire of spatio-temporal patterns of the sensory world with enough differentiation and the sensory systems as able to properly displace the system state within wide enough phase space regions, the gradual enrichment of the actual mode mixing scenario would stand for acquisition and storage of additional sensory representations in the form of specific patterns of oscillatory modes. These modes would be linked to previous ones through the mixing connections intrinsically predefined by the topological possibilities of the invariant manifolds, and could be excited again through just the same sensory way they were acquired. The incorporation of an oscillatory pattern in the mode mixing would affect an extended region of the phase space flow and, in particular, it would imply the corresponding influence upon the attractor (of the disconnected brain, in the dichotomy of our analysis). Notice however that, although externally-induced during wakefulness, the acquisition process would affect the sleeping part of the attractor, where the oscillatory activity of the asymptotic state accumulates. This peculiar fact arises inherently from the way how the mode mixing corkscrew works in the generalized Landau scenario and it is relevant for the dynamic brain because it would imply the spontaneous replay of the pattern influence on the attractor during the next sleeping stage.

Among the enormous amount of sensory input received during the wake stage of a wake-sleep cycle, those inducing oscillation modes already present would be simply confirmed; those unable to excite oscillatory traces would remain ignored; and those inducing novel oscillatory patterns would be more or less effectively acquired according to features of the previously existing memory links toward the attractor and to features of the stimuli, including the sensory context and possible repetitions. The awaking system would be always far from the endogenous asymptotic state. When asleep, some slowly varying variables regulating the sensory influence on the system, but evolving relatively independently of such an influence, would have actuated by reducing it and, leaving apart transient reminiscences of slowly decaying components, the system would evolve according to the endogenous dynamics of the sleeping part of the attractor, over which the previous waking activity would have been partly imprinted. By supposing similar plastic mechanisms during sleep, we could assume their working to adjust the system properties toward consolidation of the endogenously sustained activity actually occurring during the sleep stage. At the same time, the implicit correlation between the transient flow and the asymptotic state would imply the simultaneous consolidation of the full basin of attraction, where the transient activity of the next waking stage would take place. In this way, the sensory world presumably represented in the basin of attraction of the dynamic brain would be partially enriched during waking and fully consolidated during sleeping, every cycle, and so on in the successive wake-sleep cycles. This kind of behavior may be related to an old tradition of neurophysiology and cognitive neuroscience-based proposals for a sleep role in memory consolidation \cite{160} and particularly connected to two-stage frameworks with sleep-dependent consolidation of neuroplastic changes initiated in waking \cite{161,162}.

The feedback circuits and parameter values characterizing the (disconnected) system would continuously experience plastic changes, externally provoked during waking and endogenously sustained during sleeping, while the phase space flow describing the mode mixing of the (dynamically-relevant) variable properties could remain relatively untouched but continuously enlarged by new oscillatory patterns. In this way the changeable dynamic brain could maintain a well defined representation of the sensory world, in which the relationship between sensory items and parcels of memory would remain defined through oscillatory patterns similar to those originally induced when learning occurred. Concerning the localization of things in the dynamic brain, notice, however, that the variable properties actually experiencing the oscillatory pattern would not be the physical substrate of the corresponding memory because, as happens in any dynamical system, the dynamical evolution would arise from the whole of physical properties and their causal interrelations. Although the acquisition of a mode would happen through a given collection of spatially distributed plastic properties, successive learning and consolidation could affect the same properties again and again, and, in this way, superposed storage of memories could happen, while the oscillatory activities associated with them would affect definite sets of properties with characteristic patterns that the consolidation mechanism would try to maintain unchanged. Such a presumed tendency to maintain the previous contents of the continuously growing sensory world represented in the phase space of the dynamic brain would constitute a sort of dynamic homeostasis that, in contrast with the typical mechanisms assumed in biology on the basis of negative feedback toward a given reference status, would elude the intrinsic problem of how such a reference value should have been defined.

Let us remark the significance for the dynamic brain framework of the implicit correlation existing between the asymptotic and transient states of a basin of attraction, and also the fact that the general features of such a bidirectional correlation are mathematically not well known. Firstly, the question is relevant because it would determine the relation between the waking and sleeping cognitive experiences. For instance, in the lack of a deeper depiction, it is clear that the time evolution of the attractor would not be just a sequence of the sensory items stored in the transient flow and this could explain the illogical and bizarre features usually attributed to dreams and other sleeping experiences \cite{160}. Secondly, to what extent the transient flow/asymptotic state relation is univocal in the two senses is important because, while the waking learning processes would reflect itself in a given way on the attractor, the presumed consolidation of the sleeping part of the attractor would not need to preserve the basin of transient flow in just the reverse way and some kind of memory transformation could happen in the succession of wake-sleep cycles. Thirdly, there is the problem of how the transient flow could continuously enrich the mode mixing structure while the cyclic time of the asymptotic state remains essentially constant. The steadiness of the wake-sleep cycle could be attributed to driving influences of the day-night light cycle, but in any case it would entail some limit in the memory consolidation capacity of the dynamic brain and the influence of this limit on the memory storage capacity would depend just on the actual relation properties between the asymptotic state and its transient flow. Although the living brain ability for an indefinite permanence of stored memories remains unclear and object of debate \cite{165}, we regard this third point as a critical issue for the dynamic brain framework in the sense that could either invalidate it or indicate features of the presumed mechanisms. For instance, we find feasible an improvement of storage capacity with a given duration of the asymptotic state cycle if the mode mixing enhancement is accompanied with a dimensional growth and, to achieve this, a learning procedure with plastic influence on the definition of the dynamically relevant properties of the system would be needed.

Although the analysis has been developed by considering a learning procedure for achieving an external world representation through the sensory input, we find feasible the working of the same procedure for acquiring motor skills in the answer to external stimuli. In this case, the reduced presence of motor action during sleep should be another peculiarity to be explained through the relation between the asymptotic and transient states. We don't try to envisage the acquisition of thinking and other sequential abilities, as well as we don't want to consider what consciousness is, but the dynamic brain should reach them only through the dynamical mechanisms at its disposal.

\subsection{Summary and concluding comments}

The dynamic brain we are here conjecturing should be a single nonlinear oscillator possessing three basic merits: i) proper structure and properties for developing a generalized Landau scenario of large $(N,m)$ values, ii) proper sensory input for projecting the external world within its phase space, and iii) proper plasticity for tending to incorporate additional oscillation modes transiently induced by the sensory input. The latter feature, which would be the most meritorious, could be achieved by (discretionarily) assuming plastic mechanisms with the peculiar attribute of working as a whole to sustain positive feedback influences between the oscillatory patterns to be acquired and the induced plastic changes.
 
In interaction with its environment, the dynamic brain would effectively develop the phase space scenario in the attraction basin of one of its attractors, over which the oscillatory activity would essentially appear concentrated on a certain zone and which would provide the system with an intrinsic wake-sleep cycle. \footnote{This may be related to the ontogenetic wake-sleep cycle transformation from prenatal to adult brains \cite{166}.} During the intrinsic calm of waking, the externally provoked, but endogenously governed, transient activity would provide the system with the physical sustenance of its operative functions, including the plastic acquisition of new memories. When asleep, the system would approach the asymptotic state of the attractor and the endogenous oscillatory activity, with its slow structure of sleep phases, would sustain a service of plastic maintenance and would produce also the sleeping cognitive experiences. The peculiar relation between the asymptotic state and the transient flow of the full basin of attraction could explain the peculiar relation between the sleeping and waking experiences. \footnote{With the sleeping brain not strictly disconnected, the sleep activity would actually occur as an externally-activated transient flow in a phase space region close to the sleeping part of the attractor, provided that the sensory perturbations would be slight enough. The access to this region would be presumably feasible from within the same region only, and this could explain the commonly accepted difficulty in remembering (retrieving) dreams during waking and also the peculiar experience of volatile dream remembering after a sleep interruption.}

The dynamic brain would be a single device evolving on a single state and would therefore lack the main aspects of the {\em binding problem} \cite{168}, a problem typically found in brain models presupposing the combination of parts with differentiated operative purposes when confronted to the introspectively obvious individuality of the mind of a brain. Nevertheless, we retain that, for the reasons remarked at the end of subsection C.1, the system individuality could be hardly apparent by externally observing the detailed activity of the physical elements hypothetically sustaining the dynamic brain.

The binding problem has to do with the apparent necessity of some kind of brain controller putting order and giving sense to the working of the different components, and it is interesting to realize where the control of the dynamic brain would be, if there would be any. Since the internal management of any dynamical system lies exclusively in its structure of causal interrelations, the true control would be on the modification of this structure and, in the dynamic brain, this would imply the learning processes at two levels: first, through the topological phase space constraints on the mode mixing possibilities prefiguring features of the oscillatory patterns that a generic brain could learn and, second, through the actual history of interaction with the environment determining what a given brain could have effectively learnt. The predefinition of possibilities intrinsically associated with the generalized Landau scenario would be a sort of inherent regulatory framework that, leaving apart dimensional capabilities, could introduce commonality and homogeneity in the operative functions of different brains. For instance, the experience of pleasant (discordant) sensations in hearing music would not simply be a question of education but of proper (improper) fitting between the music-induced stream of sensory input and the predefined structures of potential oscillatory patterns. The commonality of possibilities would facilitate the composer's job, as well as the good working of any other kind of intercommunication between brains.

The dynamic brain would work by providing specific enough dynamical activities over the whole of its properties for different sensory inputs, and to do this it would have been able to construct and adjust its internal structure of dynamical relations to achieve and to maintain such a representation with enough stability. In our view, this kind of working would lack computational significance or, at least, it would have no sense to consider communication codes among the physical components of the system because the information processing would be done on the whole only. In more general terms, the dynamic brain operation could not be understood by considering a particular functional role for each of its components and by properly combining them to achieve a superior task simply because the individual roles would acquire sense within the activity of the whole only. This holistic view could be, however, compatible with some degree of spatial localization of the dynamical activities sustaining concrete operative functions, provided that the overall activity associated with the given function would occur properly oriented in the phase space.
 
Finally, in agreement with the empirical fact that it is very much easier to introduce an idea in the mind of a brain than to remove it, we find very difficult to achieve the occurrence of the opposite process of learning in the dynamic brain, i.e., the externally-induced removal of an existing oscillation mode through plastic effects of appropriate rule. We believe this reflects generic reasons of asymmetry between the building and deconstruction of oscillatory mode mixing, and this idea, together with the presumed dynamic roles of plasticity in brain development and repair, as well as the conjectured coincidence of the oscillatory scenario in both brains and turbulent fluids, has induced us to develop the framework for the structural evolution of systems presented in Sect. 4.

\end{document}